\documentclass[apj,twocolumn,twocolappendix,numberedappendix]{openjournal}
\usepackage{amsmath}
\usepackage{booktabs}
\usepackage{multirow}
\usepackage{color}
\usepackage{soul}
\usepackage{threeparttable}
\usepackage{float}
\usepackage{graphicx}
\usepackage{CJK}
\usepackage{xspace}
\usepackage{afterpage}
\usepackage{placeins}
\usepackage{amssymb}
\usepackage[breaklinks,colorlinks,citecolor=blue,urlcolor=blue,linkcolor=blue,filecolor=blue]{hyperref}
\usepackage[normalem]{ulem}
\usepackage{afterpage}

\makeatletter 
  \patchcmd{\NAT@citex}
    {\@citea\NAT@hyper@{%
      \NAT@nmfmt{\NAT@nm}%
      \hyper@natlinkbreak{\NAT@aysep\NAT@spacechar}{\@citeb\@extra@b@citeb}%
      \NAT@date}}
    {\@citea\NAT@nmfmt{\NAT@nm}%
    \NAT@aysep\NAT@spacechar\NAT@hyper@{\NAT@date}}{}{}

  \patchcmd{\NAT@citex}
    {\@citea\NAT@hyper@{%
      \NAT@nmfmt{\NAT@nm}%
      \hyper@natlinkbreak{\NAT@spacechar\NAT@@open\if*#1*\else#1\NAT@spacechar\fi}%
        {\@citeb\@extra@b@citeb}%
      \NAT@date}}
    {\@citea\NAT@nmfmt{\NAT@nm}%
    \NAT@spacechar\NAT@@open\if*#1*\else#1\NAT@spacechar\fi\NAT@hyper@{\NAT@date}}
    {}{}
\makeatother

\shorttitle{Weighing Black Hole Stars}
\shortauthors{Sun, Naidu et al.}

\usepackage[export]{adjustbox}

\newcommand{\orcidauthor}[3]{\author{\href{http://orcid.org/#1}{#2$^{#3}$}}}

\begin{document}

\title{\vspace{-1cm} Overmassive No More: The Case for Little Red Dots Hosting Black Hole Seeds\\as Massive as Single Supermassive Stars
\vspace{-1.75cm}}

\orcidauthor{0009-0007-3791-7890}{Wendy Q. Sun}{1, *}
\orcidauthor{0000-0003-3997-5705}{Rohan P. Naidu}{1, *}
\orcidauthor{0000-0003-2488-4667}{Hanpu Liu}{2}
\orcidauthor{0000-0002-2380-9801}{Anna de Graaff}{3}
\orcidauthor{0000-0002-5612-3427}{Jenny E. Greene}{2}
\orcidauthor{0000-0003-2871-127X}{Jorryt Matthee}{4}
\orcidauthor{0000-0002-5221-7557}{Chris Ashall}{1}
\orcidauthor{0000-0002-0302-2577}{John Chisholm}{5,6}
\orcidauthor{0000-0003-2895-6218}{Anna-Christina Eilers}{7}
\orcidauthor{0000-0001-7232-5355}{Qinyue Fei}{8}
\orcidauthor{0000-0002-9389-7413}{Kasper E. Heintz}{9,10}
\orcidauthor{0000-0002-1125-9187}{Daichi Hiramatsu}{11}
\orcidauthor{0000-0002-5588-9156}{Vasily Kokorev}{5,6}
\orcidauthor{0000-0001-6755-1315}{Joel Leja}{12,13,14}
\orcidauthor{0009-0002-8965-1303}{Zhaoran Liu}{7}
\orcidauthor{0000-0002-5554-8896}{Priyamvada Natarajan}{15,16,17}
\orcidauthor{0000-0001-5851-6649}{Pascal A.\ Oesch}{18,9,10}
\orcidauthor{0000-0003-3769-9559}{Robert A. Simcoe}{7}
\orcidauthor{0000-0001-5586-6950}{Alberto Torralba}{4}
\orcidauthor{0000-0001-8928-4465}{Andrea Weibel}{1}
\affiliation{$^{1}$ Institute for Astronomy, University of Hawai‘i, 2680 Woodlawn Drive, Honolulu, HI 96822, USA}
\affiliation{$^{2}$ Department of Astrophysical Sciences, Princeton University, Princeton, NJ 08544, USA}
\affiliation{$^{3}$ Max-Planck-Institut f\"ur Astronomie, K\"onigstuhl 17, D-69117 Heidelberg, Germany}
\affiliation{$^{4}$ Institute of Science and Technology Austria (ISTA), Am Campus 1, 3400 Klosterneuburg, Austria}
\affiliation{$^{5}$ Department of Astronomy, The University of Texas at Austin, Austin, TX, USA}
\affiliation{$^{6}$ Cosmic Frontier Center, The University of Texas at Austin, Austin, TX 78712, USA}
\affiliation{$^{7}$ MIT Kavli Institute for Astrophysics and Space Research, 70 Vassar Street, Cambridge, MA 02139, USA}
\affiliation{$^{8}$ David A. Dunlap Department of Astronomy \& Astrophysics, University of Toronto, 50 St. George St., Toronto, ON M5S 3H4, Canada}
\affiliation{$^{9}$ Cosmic Dawn Center (DAWN), Copenhagen, Denmark}
\affiliation{$^{10}$ Niels Bohr Institute, University of Copenhagen, Jagtvej 128, K{\o}benhavn N, DK-2200, Denmark}
\affiliation{$^{11}$ Department of Astronomy, University of Florida, Bryant Space Science Center, Gainesville, FL 32611-2055 USA}
\affiliation{$^{12}$ Department of Astronomy \& Astrophysics, The Pennsylvania State University, University Park, PA 16802, USA}
\affiliation{$^{13}$ Institute for Gravitation and the Cosmos, The Pennsylvania State University, University Park, PA 16802, USA}
\affiliation{$^{14}$ Institute for Computational \& Data Sciences, The Pennsylvania State University, University Park, PA 16802, USA}
\affiliation{$^{15}$ Department of Astronomy, Yale University, New Haven, CT 06511, USA}
\affiliation{$^{16}$ Department of Physics, Yale University, New Haven, CT 06511, USA}
\affiliation{$^{17}$ Yale Center for Astronomy \& Astrophysics, Yale University, New Haven, CT 06520, USA}
\affiliation{$^{18}$ Department of Astronomy, University of Geneva, Chemin Pegasi 51, 1290 Versoix, Switzerland}

\thanks{$^*$E-mail: \href{mailto:qinyi.sun@hawaii.edu}{qinyi.sun@hawaii.edu}, \href{mailto:rohan.naidu@hawaii.edu}{rohan.naidu@hawaii.edu}}

\begin{abstract}
Little Red Dots (LRDs) display singular properties unlike any known class of AGN or galaxies, motivating novel mass estimators for their central engines. Inspired by their similarities to stellar phenomena, here we interpret the LRD continuum as being produced by a pseudo-photosphere. We fit tailored stellar atmosphere models to host-subtracted LRD central engines (``black hole stars,'' BH*s) represented by stacks of $117$ objects. Typical BH* continuum spectra are well fit by models in a narrow range of temperatures ($T_{\rm eff}\approx4200-4800$ K), with bolometric luminosities $\approx10^{43-45}$ erg s$^{-1}$, implying pseudo-photospheric radii $\approx700-2000$ au. Based on these parameters, we explore four different approaches to deriving BH* masses: 1) using the surface gravity from atmosphere models; 2) appealing to the resemblance to super-Eddington phenomena; 3) approximating the escape velocity from the outflowing material; and 4) exploiting the lack of variability to bound the dynamical time. For the typical BH*, all of these methods yield remarkably consistent masses of $\approx10^{4-5}\,M_\odot$, implying a highly super-Eddington luminosity of $L_{\rm{bol}}/L_{\rm{Edd}}\sim5-50$. These mass estimates place BH*s within the scatter of the local scaling relation between black hole mass and host galaxy stellar mass, providing a self-consistent alternative to ``overmassive'' black holes that lie $2-3$ dex above it. Crucially, our derived masses are consistent with BH*s arising from single supermassive stars (SMSs), whose masses cannot exceed $\approx10^{5-6}\,M_\odot$ due to general relativistic instabilities. Furthermore, for our derived $L_{\rm bol}/L_{\rm Edd}$, the sharp cutoff of the LRD luminosity function matches the maximum theoretical mass of an SMS. With LRDs, we may therefore be directly observing the birth of heavy black hole seeds.
\end{abstract}

\section{Introduction}
\label{sec:intro}

Little Red Dots \citep[LRDs;][]{Matthee24} are red, compact sources discovered by the James Webb Space Telescope (JWST), characterized by ``V-shaped'' spectral energy distributions (SEDs), point-like morphology, and broad Balmer emission lines \citep[e.g.,][]{Greene24, Kocevski24, Kokorev24, Akins24, Hviding25}. Found in almost every relatively deep JWST image, hundreds of LRDs have now been reported across $z\approx0-10$ \citep[e.g.,][]{Lin25, Park26, Kapoor26, Zhang25GNLRDs, Tanaka25}. Remarkably, despite three years of intense attention, the most basic of their parameters -- the masses of their central engines\footnote{In detail, the ``central'' engines (BH*s) in LRDs may lie far from the centers of their hosts \citep[e.g.,][]{Yanagisawa26}.} -- remains debated by several orders of magnitude \citep[e.g.,][]{Juodzbalis25direct, Naidu26BHstar, Greene25, Chang26, Rusakov26}, spanning the regime from seed black holes ($\approx10^{3-5} M_{\odot}$; e.g., \citealt{Umeda25, Liu26TLUSTY, Gentile26, Yanagisawa26, Naidu26}) to supermassive black holes ($\approx10^{6-8} M_\odot$; e.g., \citealt{Taylor24, Jones25, Maiolino25, Gupta26, Kocevski26}). 

The difficulty in determining the central engine mass is that the workhorse method for such measurements -- virial estimators based on broad Balmer lines \citep[e.g.,][]{GreeneHo05, Reines13} -- may not apply to LRDs. LRD Balmer lines find no peer among classical AGN used to calibrate virial relations. They display a \textit{simultaneous combination} of extreme EWs \citep[e.g.,][]{Sun26}, Balmer absorption \citep[e.g.,][]{Yanagisawa26c}, strong deviations from Case-B recombination ratios \citep[e.g.,][]{Lin25}, lack of variability \citep[e.g.,][]{Liu26TWINKLE}, and electron-scattered wings \citep[e.g.,][]{Rusakov26}. Furthermore, LRDs are faint in X-rays \citep[e.g.,][]{Yue24, Ananna24, Sacchi25}, lack hot and cold dust emission \citep[e.g.,][]{Casey25, Xiao25, Setton25}, and typically display little optical continuum variability on $\approx10$ yr timescales \citep[e.g.,][]{Kokubo25, Tee25, Zhang25var, Burke26, Liu26TWINKLE}. When applied to LRDs, virial estimators calibrated on local AGN yield black hole masses of $\sim 10^{7-8} \, M_\odot$ \citep[e.g.,][]{Chen25hostifany, Jones25, Maiolino25}, which sit orders of magnitude above the $z=0$ scaling relation between black hole mass and stellar mass \citep[e.g.,][]{RV15}. More strikingly, in some cases the black hole masses almost exceed the stellar masses of the host galaxies, thus implying extremely ``overmassive'' black holes when compared to the local $M_{\rm{BH}}/M_{\rm{\star}}<0.1\%$ \citep[e.g.,][]{Juodzbalis25direct, deugenio25twice, Ivey26}. If the typical LRD is indeed ``overmassive,'' given their ubiquity we are faced with the conundrum that the BH mass density at $z\approx5$ already rivals that of the local Universe \citep[e.g.,][]{Luberto25}.

Nor can direct dynamical measurements of the central engine mass be easily obtained. Direct measurements rely on resolving gas or stellar motions within the black hole sphere of influence to infer the enclosed mass from Keplerian rotation. The sphere of influence of a black hole is $r_{\rm{infl}} = GM_{\rm{BH}}/\sigma^2$, where $\sigma$ is the velocity dispersion of the host. Narrow forbidden lines in LRDs indicate $\sigma \approx 50 - 100 \, \rm{km} \, \rm{s}^{-1}$ \citep[e.g.,][]{Ji25BT, DEugenio26z5lrd, Matthee26}. For $\sigma \approx 50 \, \rm{km} \, \rm{s}^{-1}$, a $\approx 10^7 \, M_\odot$ black hole at $z \approx 5$ has $r_{\rm{infl}} \approx 20$ pc, subtending $\approx 5$ mas, more than an order of magnitude lower than the diffraction limit of the JWST NIRSpec/IFU at the observed wavelengths of H$\alpha$ and H$\beta$. Even a dynamically cold host with $\sigma \approx 15 \, \rm{km} \, \rm{s}^{-1}$ yields only $\approx 30$ mas. Strong lensing may help occasionally bridge this gap \citep[e.g.,][]{Juodzbalis25direct}, in which case a point mass must be distinguished from an extended nuclear star cluster.

Traditional calibrations may not hold and direct measurements are largely out of reach, but what must replace them is as yet unclear. Against this backdrop, in this paper we explore emerging approaches to weighing LRD central engines. A key insight is that the central engines of LRDs may be black holes enshrouded in dense cocoons of gas such that they radiate in a manner reminiscent of stellar phenomena -- i.e., they are ``black hole stars'' \citep[BH*s;][]{Naidu26BHstar, degraaff25, degraaff25pop}. The name captures a duality. LRDs radiate at luminosities on the scale traditionally associated with black holes \citep[e.g.,][]{Greene25}, yet display signatures classically associated with stellar phenomena such as Balmer breaks \citep[e.g.,][]{Setton24}, blackbody-like continua \citep[e.g.,][]{degraaff25pop}, and photospheric absorption features \citep[e.g.,][]{Lin25}. Initially, this analogy was drawn to true photospheres, with the continuum arising from an envelope paralleling a stellar atmosphere \citep[e.g.,][]{Kido25, Begelman25, Inayoshi25coevol}. Recently, \citet[][]{Naidu26} further extended the analogy from quasi-static envelopes to dynamic outflows, noting numerous parallels between LRDs and massive star eruptions that are trapped within dense circumstellar material (e.g., Type IIn supernovae, great eruptions of luminous blue variables, luminous red novae). In this picture, the blackbody-like continuum arises from a pseudo-photosphere formed within the outflow, self-regulated by hydrogen recombination opacity to $T_{\rm{eff}} \approx 4000-7000$ K \citep[e.g.,][]{Rest12, Prieto14, Smith18, degraaff25pop, Umeda25, Sun26}. It is accompanied by pseudo-photospheric absorption lines \citep[Ca H\&K, Calcium Triplet, Na D;][]{deugenio25irony, Lin26, Liu26TLUSTY} and molecular tracers from cooler layers that extend to large radii along a density/temperature gradient \citep[e.g.,][]{Prieto14, Wang26}. The outflow itself imprints P-Cygni profiles seen clearly in e.g. \ion{He}{1}, H$\alpha$, and H$\beta$ \citep[e.g.,][]{Smith18, NM24, Juodzbalis24rosetta, Loiacono25, Matthee26}. Electron scattering in the partially ionized wind produces broad exponential wings \citep[e.g.,][]{Dessart09, Smith18, Torralba25IFU, Rusakov26, Sneppen26}, while collisional excitation and Lyman pumping in the same wind generate an \ion{Fe}{2} forest and strong \ion{O}{1} \citep[e.g.,][]{Smith18, deugenio25irony, Tripodi25, Kokorev25glimpsed, Torralba25IFU, Lin26, degraaff25pop}. If these parallels between LRDs and various stellar phenomena are physical, they suggest alternative mass estimators, which we explore here.

Regardless of which quasi-stellar picture one adopts, the BH* masses that follow are likely lower than virial estimators suggest. First, since much of the broad Balmer line width arises from electron scattering rather than virial motion (provided a kinematic broad-line region even exists in the first place) \citep[e.g.,][]{Chang26, Sneppen26, Rusakov26}, the intrinsic lines are considerably narrower, resulting in correspondingly lower masses. Second, in the outflow picture, there may be no virial intrinsic line width to recover at all, with the lines forming instead entirely in a wind \citep[e.g.,][]{Naidu26, Chisholm26, Martins26}, whose widths trace the velocity of the outflow rather than virial motion. Third, the bolometric luminosity is likely far lower than standard AGN bolometric corrections would suggest, since LRDs have far weaker dust emission \citep[e.g.,][]{Casey25, Setton25, Xiao25} and X-rays \citep[e.g.,][]{Yue24, Ananna24, Sacchi25} compared to classical AGN. For an engine radiating near the Eddington limit, a lower luminosity implies a lower mass. Together, these considerations mean that any quasi-stellar picture, once adopted, implies masses lower than the virial estimates.

\begin{figure*}
    \centering
    \includegraphics[width=\linewidth]{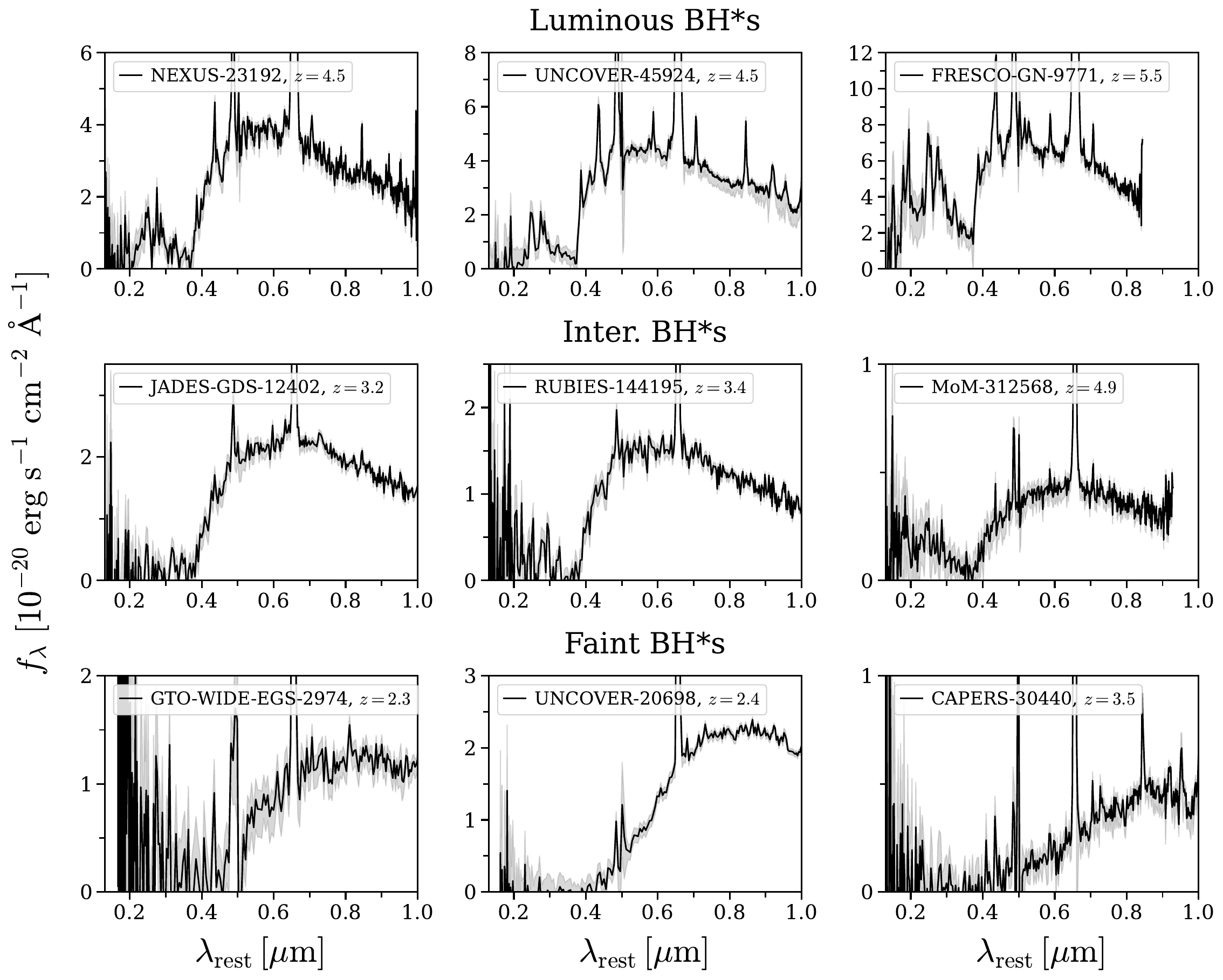}
    \caption{\textbf{Examples of host-subtracted LRD central engines (BH*s for short) in three substacks split by $L_{5500}$ demonstrate spectral differences.} The luminous BH*s (top) are defined to have $44.0 < \log\left(L_{\rm{5500}} / \text{erg}\,\text{s}^{-1}\right) < 45.0$; the intermediate BH*s (middle) have $43.0 < \log\left(L_{\rm{5500}} / \text{erg}\,\text{s}^{-1}\right) < 44.0$; the faint BH*s (bottom) have $42.0 < \log\left(L_{\rm{5500}} / \text{erg}\,\text{s}^{-1}\right) < 43.0$. Notice the diversity in spectral shape that falls out of this luminosity separation -- e.g., luminous BH*s peak at bluer wavelengths and are UV-bright, whereas faint BH*s peak at redder wavelengths and have little UV flux \citep[e.g.,][]{Zhang25, Ando26, Cloonan26}. Note that the high SNR of the selected examples here is not representative of the entire sample of objects we use.}
\label{fig:substack_gallery}
\end{figure*}

\begin{figure*}
    \centering
    \includegraphics[width=\linewidth]{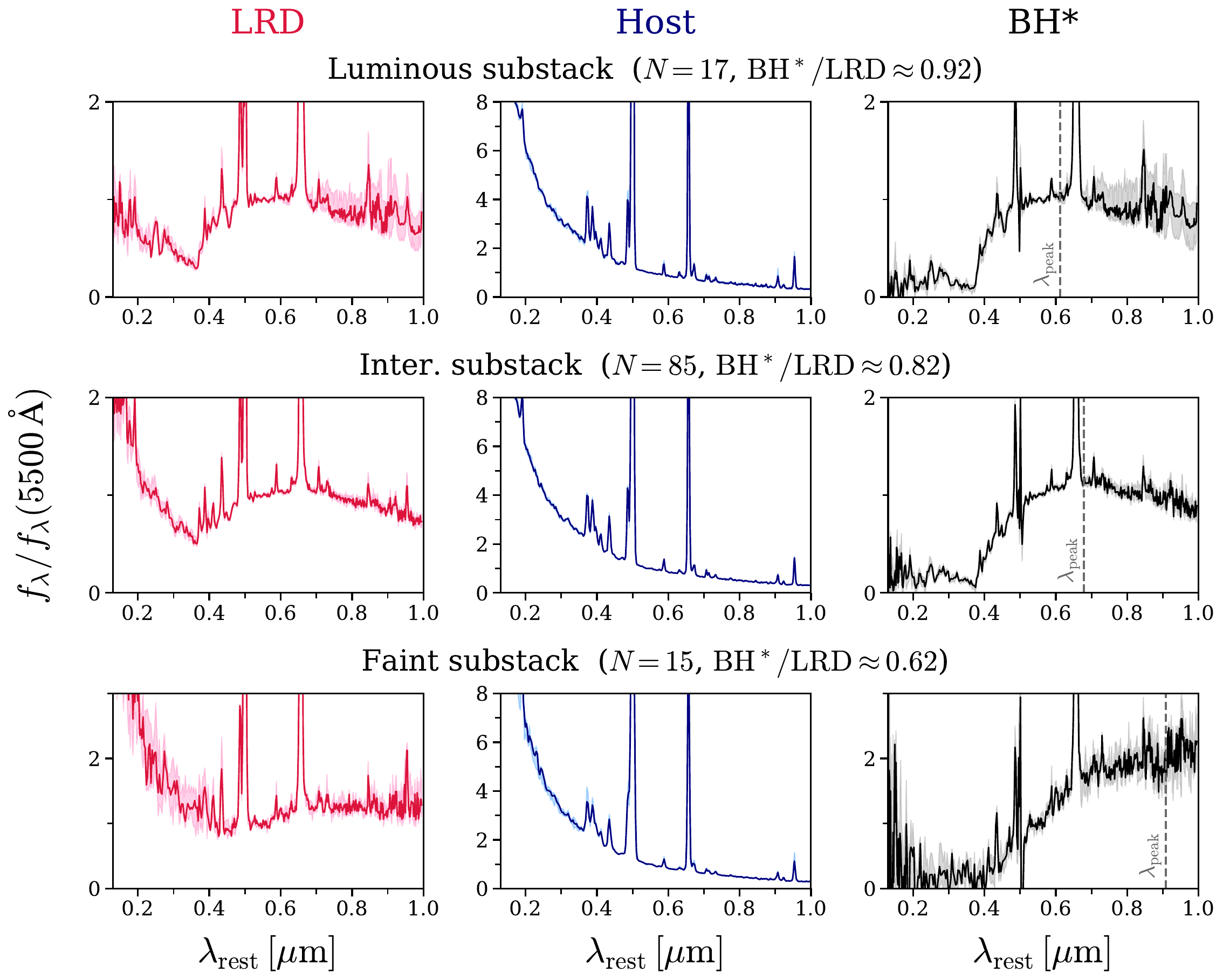}
    \caption{\textbf{LRD $-$ host $=$ BH* decomposition for the three substacks.} Following the decomposition procedure in \cite{Sun26}, we assume that [\ion{O}{3}]5008\AA\ arises exclusively from the host galaxy. Each LRD is matched to galaxies at similar redshift and [\ion{O}{3}] luminosity; each candidate host is rescaled to account for all of the [\ion{O}{3}] before subtracting. The BH* spectra therefore have zero [\ion{O}{3}] by construction, and any residual at that wavelength reflects small differences in line width between the LRD and its candidate hosts. The median stacks are constructed via bootstrap resampling, and the shaded bands indicate uncertainties on the median from bootstrapping (16$^{\rm{th}}$ and 84$^{\rm{th}}$ percentiles). The median BH*/LRD fraction is evaluated at 5500\AA. The Balmer break and red optical continuum characteristic of BH*s emerge clearly after subtraction. The more luminous a BH*, the bluer its peak wavelength -- we mark the peak wavelength $\lambda_{\rm{peak}}$ obtained from modified blackbody fits ($\lambda_{\rm{peak}} = 6122, \, 6792, \, 9081 \rm{\AA}$ for the luminous, intermediate, and faint BH*s, respectively).}
\label{fig:substacks}
\end{figure*}

Settling the mass debate is critical, as the BH* masses could discriminate between supermassive black hole (SMBH) formation channels. The difference between $\sim 10^{3-5} \, M_\odot$ and $\sim 10^{6-8} \, M_\odot$ is the difference between witnessing heavy seeds near their formation and finding SMBHs already assembled by $z \sim 5$. One channel that the massive star eruption-like character of LRDs invites us to test is: Black Hole Stars may form from single supermassive stars (SMSs; $10^{3-5} \, M_\odot$) observed in a state similar to $\eta$ Carinae during its Great Eruption \citep[e.g.,][]{Nandal26smslrds, Nandal26, Zwick26, Chisholm26, Martins26, Naidu26}. Such SMSs are expected to collapse into massive black hole seeds. A basic challenge to the SMS scenario is that an SMS has a maximum mass of $\approx 10^{5} \, M_\odot$ due to general relativistic instabilities \citep[e.g.,][]{Chandrasekhar64, Fowler66, Woods17}, with optimistic assumptions (e.g., extreme accretion rates) allowing for only slightly larger masses of $\approx10^{6} \, M_\odot$ \citep[e.g.,][]{Nandal24gr, Saio24gr}. Virtually all existing mass estimates, then, at face value, rule out the scenario in which BH*s arise from single SMSs.

Our goal in this paper is twofold: we will explore alternative paths to BH* masses that are independent of virial calibrations, and then as an application, ask whether the resulting masses open the possibility of BH*s arising from single SMSs. This paper is organized as follows. In \S\ref{sec:methods}, we describe how we use the atmosphere models from \cite{Liu26TLUSTY} to obtain pseudo-photospheric parameters required for mass estimates. Then, in \S\ref{sec:results}, we present results on the derived BH* masses. Finally, in \S\ref{sec:discussion}, we discuss the implications for an SMS origin of BH*s, predict observational signatures, and examine the scope over which the mass estimates apply. Throughout this work, we adopt a flat $\Lambda$CDM cosmology with parameters as per \citet[][]{Planck18}.

\section{Methods}
\label{sec:methods}

To weigh the BH*s without contamination from their host galaxies, we first subtract the hosts to isolate the BH* spectra. We then fit optically thick atmosphere models from \cite{Liu26TLUSTY} to the host-subtracted BH* stacks to estimate their ``fundamental stellar parameters.'' The model fitting outputs a self-consistent set of parameters, including effective temperature $T_{\rm{eff}}$, bolometric luminosity $L_{\rm{bol}}$, surface gravity $\log{g}$, metallicity [M/H], and microturbulence velocity $\xi_{\rm{mtb}}$. We then mainly use the effective pseudo-photospheric radius $R_{\rm{phot}}$ implied by the Stefan-Boltzmann law $L_{\rm{bol}}= 4\pi \sigma_{\rm{SB}} R_{\rm{phot}}^{2} T_{\rm{eff}}^{4}$ to estimate BH* masses.

\subsection{LRD $-$ Host Galaxy $=$ BH* Decomposition}
\label{sec:lrd_decomposition}

We disentangle the central engines of LRDs from their host galaxies following \cite{Sun26}, who assume that the [\ion{O}{3}]5008\AA\ line, on average, arises exclusively from the host galaxy -- motivated by its narrow width in LRDs, its tight correlation with $M_{\rm{UV}}$, and its near-absence in the two pure BH*s MoM-BH*-1 and the Cliff. Each LRD is matched to galaxies from the public DAWN JWST Archive\footnote{\url{https://dawn-cph.github.io/dja/index.html}} \citep[DJA;][]{DJA, Heintz25} at similar redshift and [\ion{O}{3}] luminosity; each candidate host galaxy is rescaled to account for all of the [\ion{O}{3}] and subtracted, generating a posterior of BH* spectra per source. \cite{Sun26} validate this pipeline with end-to-end mock tests, which recover continuum luminosities, Balmer break strengths, and line EWs to $\approx 10\%$, with residuals within $\sim 1\sigma$ at $\lambda_{\rm{rest}} \gtrsim 2,000\rm{\AA}$. The host-subtracted spectra display the hallmark features of BH*s -- including a blackbody-like optical continuum, extreme H$\alpha$ EWs, steep Balmer decrements, a Balmer break far exceeding what any stellar population can produce, and \ion{Fe}{2} and \ion{O}{1} emission from dense gas. These features motivate the pseudo-photospheric interpretation, on which the rest of this paper is conditioned.

Here, we apply the decomposition procedure to an expanded LRD sample based on updates to \citet{degraaff25pop} using \texttt{v4.5} of DJA \citep[][]{DJA, Valentino25, Pollock26}. As before, we also include five LRDs discovered with the NIRCam grism in the FRESCO Survey \citep[][]{Oesch23, Matthee24}, which were followed up with NIRSpec/IFU (PRISM+G395H) in GO-5664 (PI: Matthee). After removing duplicates and performing quality cuts as described in \cite{Sun26}, we obtain median BH* spectra for $117$ LRDs.

Individual BH* spectra are generally too noisy to yield useful constraints, so we adopt a stacking approach. We split the $117$ BH*s into three bins based on their $L_{5500}$ -- luminous ($44.0 < \log\left(L_{\rm{5500}} / \text{erg}\,\text{s}^{-1}\right) < 45.0$, $N=17$), intermediate ($43.0 < \log\left(L_{\rm{5500}} / \text{erg}\,\text{s}^{-1}\right) < 44.0$, $N=85$), and faint ($42.0 < \log\left(L_{\rm{5500}} / \text{erg}\,\text{s}^{-1}\right) < 43.0$, $N=15$). Fig. \ref{fig:substack_gallery} shows example BH*s in each bin. Then, we construct three corresponding median substacks in a Monte Carlo fashion across $10,000$ trials by taking the median BH* spectrum for each of the LRDs in a bin, bootstrap resampling these spectra, and computing the median spectrum for each trial. We report the median across trials, with the 16$^{\rm{th}}$ and 84$^{\rm{th}}$ percentiles as the uncertainties. The decomposition for the three substacks is shown in Fig. \ref{fig:substacks}.

The BH* substacks split by luminosity reveal a clear diversity in spectral shape. The luminous BH*s peak at bluer rest-frame wavelengths than the faint BH*s \citep[e.g.,][]{degraaff25pop}, as shown by the modified blackbody peak wavelengths marked in Fig. \ref{fig:substacks}. Also, the luminous BH*s are UV-bright, whereas the faint ones have little UV flux \citep[e.g.,][]{Zhang25, Ando26, Cloonan26}. Given this spectral shape diversity, we fit each BH* substack separately in addition to the median combined stack, instead of only the combined stack.

\subsection{Atmosphere Models}
\label{sec:models}

\cite{Liu26TLUSTY} computed a synthetic spectral library of optically thick atmosphere models tailored to LRDs, using \texttt{TLUSTY} \citep[][]{Hubeny88, Hubeny21} for the atmosphere structure and \texttt{SYNSPEC} \citep[][]{Hubeny11, Hubeny21} for spectral synthesis. It may seem somewhat surprising that we are deploying stellar atmosphere models to describe what may be possibly a dynamic pseudo-photosphere. However, such models are routinely used in e.g. the Great Eruption literature \citep[e.g.,][]{Rest12, Prieto14, Smith18} as more robust estimators of e.g. the effective temperature than simple blackbody fits \citep[e.g.,][]{degraaff25pop, Sun26, Umeda25}, since they account for the covariance of temperature, density, and metallicity simultaneously. It is primarily for this purpose that we deploy these models.

We use the models here for fitting the BH* pseudo-photosphere continuum, without accounting for the layers that imprint features such as the sharp Balmer break, emission lines, and Balmer absorption. As such, these are models of the pseudo-photosphere continua, and not of the total BH* emission \citep[cf.][]{Juodzbalis26}. The models assume LTE and plane-parallel geometry, and include line lists of atoms, ions, diatomic molecules, and water. The grid spans $2000 \, {\rm K} \leq T_{\rm{eff}} \leq 7500 \, \rm{K}$ and $-4 \leq \log{g} \leq 1.5$, extending to much lower gravities than standard stellar libraries, with the densest coverage at $T_{\rm{eff}} = 4500 - 5000 \, \rm{K}$ and [M/H] $= -1.0$. As discussed earlier, the key parameters of interest we constrain using these models are $T_{\rm{eff}}$ and $L_{\rm{bol}}$ for $R_{\rm{phot}}$ estimates.

Additionally, the fits also produce estimates for the net gravity $g_{\rm{net}} = g - g_{\rm{dyn}}$, where $g$ is the true surface gravity and $g_{\rm{dyn}}$ accounts for dynamical support from gas motion. Radiative transfer never directly sees $g$, which enters through the photospheric density structure by hydrostatic equilibrium, so the spectrum constrains the net gravity $g_{\rm{net}}$. In other words, atmospheres with the same effective temperature and photospheric density give similar spectra whether the acceleration is gravitational or dynamical in nature. The models can constrain $g_{\rm{net}}$ because the emergent spectrum depends on it in several distinct ways. \cite{Liu26TLUSTY} show that SEDs of optically thick atmospheres may appear narrower or broader than simple blackbodies due to $g_{\rm{net}}$, and that the H$^-$ kink and the CaT absorption strength are also sensitive. In this paper, we are fitting the optical-to-near-IR continuum shape, which we find produces compelling constraints. We fit $g_{\rm{net}}$ throughout; hereafter, we simply call it $g$.

Low-gravity atmospheres can become super-Eddington (i.e., $g < g_{\rm{rad}}$, where $g_{\rm{rad}}$ is the radiative acceleration), and \cite{Liu26TLUSTY} treat two regimes accordingly. In the first regime (``hydrostatic'', $220$ out of $233$ models), only the inner layers become super-Eddington and drive gas pressure inversion. Because such inversion is physically unstable, those inner layers are held at the maximum gas pressure, leaving the outer atmosphere (where spectral features form) hydrostatic; $g_{\rm{rad}}$ is otherwise self-consistently considered. In the second regime with even lower gravity (``no-$g_{\rm{rad}}$'', $13$ out of $233$ models), the entire atmosphere becomes super-Eddington, and $g_{\rm{rad}}$ is dropped. These are not physical hydrostatic structures but a prescription for an optically thick gas layer at a very low density, and are computed only at $T_{\rm{eff}} = 4500, \, 5000 \, \rm{K}$ with fixed [M/H] $= -1.0$ and $\xi_{\rm{mtb}} = 2 \, \rm{km \, s}^{-1}$. We fit the hydrostatic models throughout, as they are physically self-consistent and less affected by uncertainties in NLTE and geometry.

\begin{deluxetable}{llc}
\tablecaption{Atmosphere Model Parameters and Priors\label{tab:model_priors}}
\tablehead{
    & \multicolumn{1}{l}{Parameter} & \colhead{Prior}
}
\startdata
Effective temperature & $T_{\rm eff}$ (K) & $[2000, 7500]$ \\
Surface gravity & $\log(g/{\rm cm\,s^{-2}})$ & $[-4, 1.5]$ \\
Metallicity & $[{\rm M/H}]$ & $[-2, 0]$ \\
Microturbulence velocity & $\xi_{\rm mtb}$ (km\,s$^{-1}$) & $[2, 10]$ \\
Uncertainty parameter & $\log{s}$ & $[-2, 1]$ \\
Scale parameter & $\log{\texttt{norm}}$ & $[-1, 1]$
\enddata
\tablecomments{Fits use the hydrostatic model grid.}
\end{deluxetable}

\begin{figure*}[p]
    \centering
    \includegraphics[width=\linewidth]{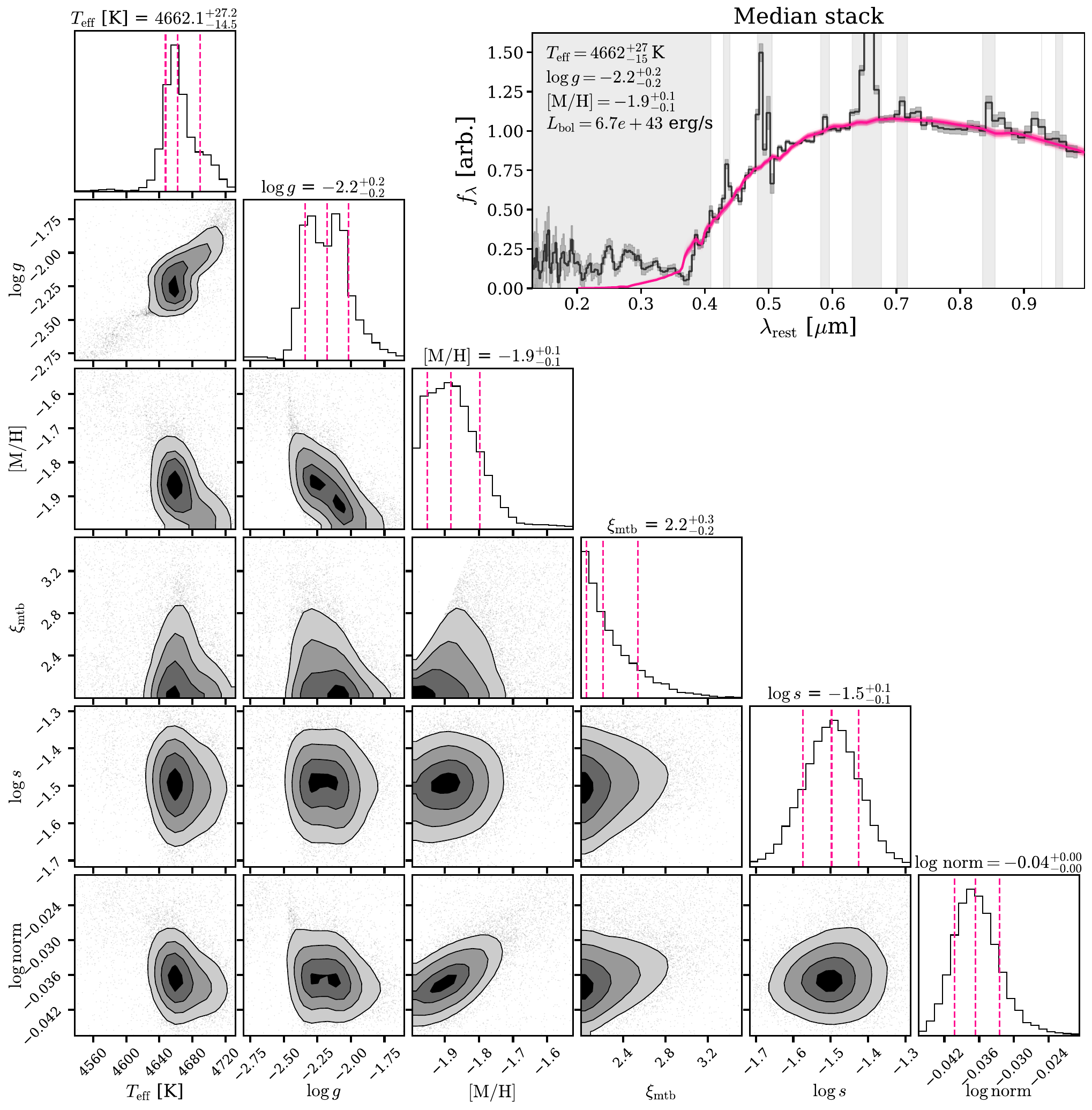}
    \caption{\textbf{Well-constrained pseudo-photospheric parameters from \texttt{dynesty} posterior for the median BH* stack.} We adopt uniform priors for all parameters, as listed in Table \ref{tab:model_priors}. Marginal and joint distributions are shown for the six parameters $\left( T_{\rm{eff}}, \, \log{g}, \, \text{[M/H]}, \, \xi_{\rm{mtb}}, \, s, \, \texttt{norm} \right)$; contours and histograms are importance-weighted over the full set of nested samples. Dashed lines and row titles report the median and 16$^{\rm{th}}$/84$^{\rm{th}}$ percentiles of the marginal posterior for each parameter. The [M/H] and $\xi_{\rm{mtb}}$ posteriors are concentrated at the lower prior bounds; they remain meaningfully constrained, though limited by the extent of the model grid. (Top right) \textbf{Atmosphere model fits for the median BH* stack} (black), where $1{,}000$ equal-weight posterior draws are plotted (pink), with the per-wavelength median of the draws as the thick curve. Shaded vertical bands mark wavelengths excluded from the fit, including strong emission lines and $\lambda_{\rm{rest}} \leq 4100\rm{\AA}$, so that the fit constrains the continuum shape redward of the Balmer break.}
\label{fig:atm_fits_stack}
\end{figure*}

\afterpage{
\onecolumngrid
\begin{center}
    \includegraphics[width=\linewidth]{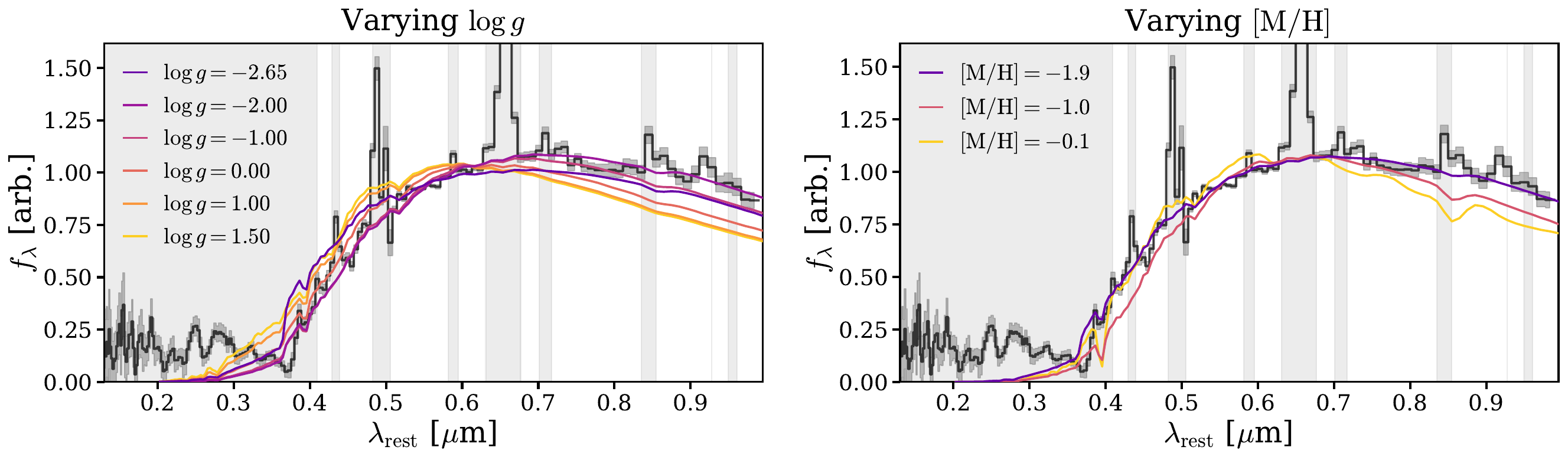}
    \figcaption{\textbf{Pseudo-photosphere continuum shape of the models is sensitive to $\log{g}$ and [M/H].} The colored curves show models in which one parameter is swept while $T_{\rm{eff}}$, $\xi_{\rm{mtb}}$, the other of $\log{g}$/[M/H], and the flux normalization are all fixed at their marginal posterior medians. Shaded vertical bands mark wavelengths excluded from the fit. The model spectra are sensitive to both parameters, which is what allows the fitting to constrain them. For $\log{g}$, the near-IR H$^-$ opacity scales more steeply with gas density than the optical metal-line opacity. For [M/H], metallicity impacts the SED through various mechanisms, e.g., by changing metal absorption opacities and by contributing free electrons for H$^-$ formation that dominates the near-IR opacity at these temperatures.\label{fig:vary_parameter}}
\end{center}
\twocolumngrid
}

Before proceeding further, we list some of the limitations inherent in these models. The atmosphere models are necessarily a simplification. We emphasize that although the LRD structure is likely not hydrostatic (with e.g. outflows, as evidenced by the P-Cygni line profiles), the hydrostatic models can still provide a useful description that yields self-consistent parameters. Assuming LTE, the models ignore an array of NLTE effects \citep[e.g.,][]{Martins26} such as collisional excitation, fluorescence, and pumping that are thought to produce several rest-optical emission lines \citep[e.g.,][]{Tripodi25, Torralba25IFU, Yan25, Kokorev25glimpsed, Chang26}. Separately, the hydrostatic models assume no significant gas content far from the continuum-forming pseudo-photosphere, whereas in reality, such gas layers with well-populated $n=2$ hydrogen levels are likely responsible for producing the observed sharp Balmer break, Balmer emission, and Balmer absorption. The gas is also likely inhomogeneous rather than confined to a smooth layer -- e.g., in the form of clumps and bullets as seen in luminous blue variable (LBV) eruptions such as the \cite{Weigelt86} blobs. Overall, the emission lines and absorption features are very sensitive to the detailed structure of such gas layers, and we do not model them. Our fitting is based on the assumption that the optical-to-near-IR continuum, with a relatively low opacity, primarily depends on gas conditions at the pseudo-photosphere, which may be represented by parameters from the hydrostatic atmosphere models. As we will describe below, these models produce excellent fits to the shape of the BH* continuum.

\subsection{Atmosphere Model Fitting}
\label{sec:model_fitting}

The model spectra are convolved with the NIRSpec/PRISM dispersion curve to match the instrumental resolution of the data, rebinned onto $200$ logarithmically equally spaced wavelength bins from $\approx 1000 - 10000\rm{\AA}$ (the wavelength range of the stacks), and normalized by the continuum median calculated over all unmasked windows. We linearly interpolate the resulting library over $T_{\rm{eff}}$, $\log{g}$, [M/H], and $\xi_{\rm{mtb}}$. As the grid does not form a complete rectangular product (i.e., not all parameter combinations occur over the full range of each parameter), we linearly interpolate over its Delaunay tessellation rather than by tensor product \citep[][]{scipy}, rescaling each parameter to unit range before tessellating. Proposals falling outside of the convex hull of the grid are assigned zero likelihood, limiting the effective prior volume to that captured by the grid. We adopt uniform priors for all parameters, as listed in Table \ref{tab:model_priors}. The free parameter $s$ accounts for unknown systematic uncertainty in the model, and the scale parameter \texttt{norm} allows the model flux level to match the BH* stack since only the spectral shape is informative.

\begin{figure}
    \centering
    \includegraphics[width=\linewidth]{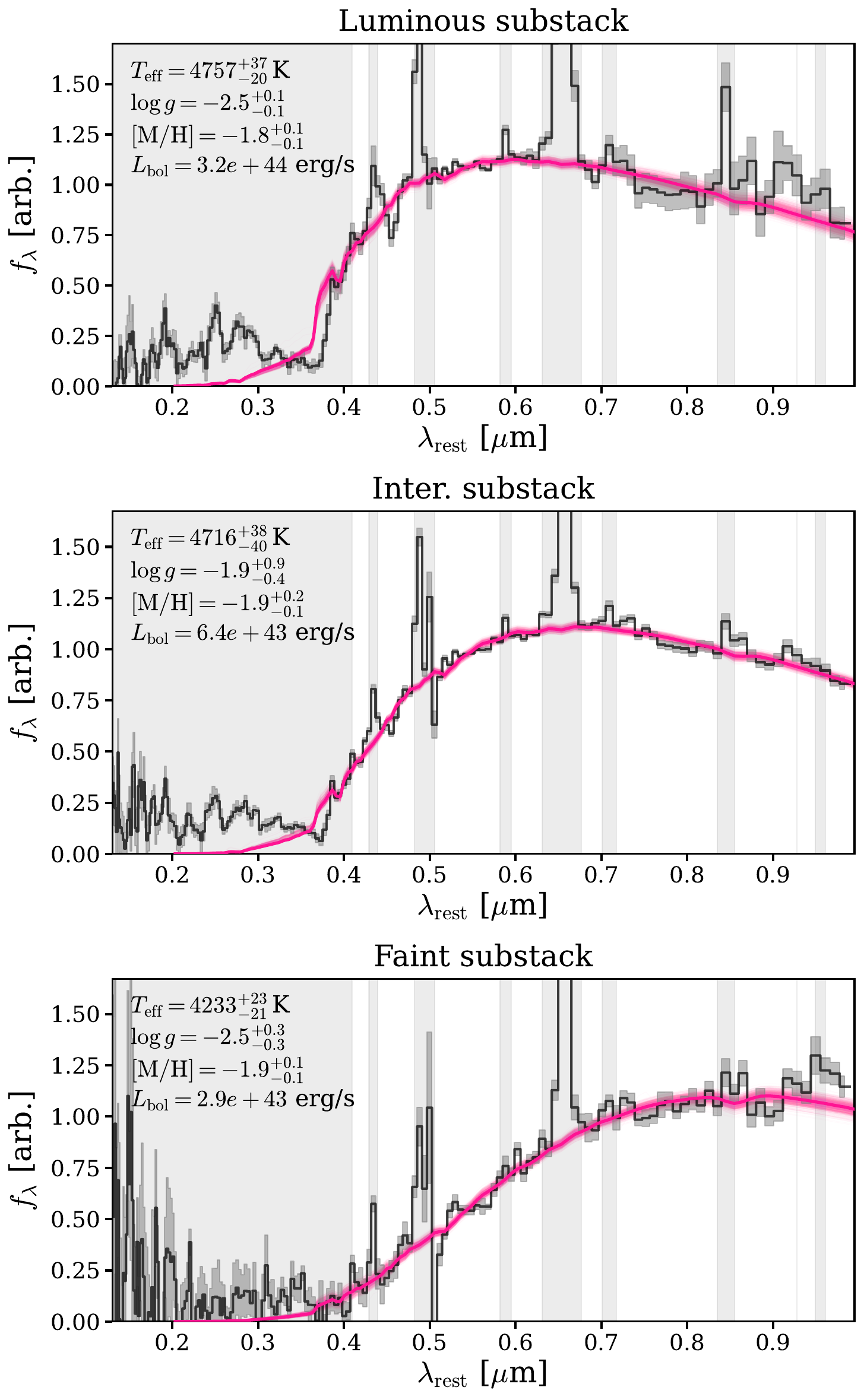}
    \caption{\textbf{Atmosphere model fits for the three BH* substacks} (black), where $1{,}000$ equal-weight posterior draws are plotted (pink), with the per-wavelength median of the draws as the thick curve. Shaded vertical bands mark wavelengths excluded from the fit. Following Fig. \ref{fig:atm_fits_stack}.}
\label{fig:atm_fits_substacks}
\end{figure}

Similarly, each BH* stack is rebinned onto $200$ logarithmically equally spaced wavelength bins and normalized by its continuum median. We fit only $\lambda_{\rm{rest}} > 4100 \rm{\AA}$, masking the region blueward of the Balmer break as well as strong emission lines. Note that the models represent the pseudo-photosphere continuum \citep[e.g.,][]{Naidu26} -- hence, features such as the sharp Balmer breaks or deep Balmer absorption of BH*s are not captured in the models. Our fit effectively focuses on the optical-to-near-IR continuum shape. Following \cite{Liu26TLUSTY}, we adopt the following logarithmic likelihood to evaluate a model against the BH* spectrum,
\begin{align*}
    \ln{P} \, = \, -\frac{1}{2} \, \sum_{\text{bin}} &\Biggl[\, \frac{\left(f_{\lambda,\rm{stack}} - f_{\lambda,\rm{model}}\right)^2}{\sigma^2_{\lambda,\rm{stack}} + s^2 f^2_{\lambda,\rm{model}}} \\
    &\quad + \ln\left(\sigma^2_{\lambda,\rm{stack}} + s^2 f^2_{\lambda,\rm{model}}\right) \, \Biggr] \, \begin{matrix} \\,\\ \end{matrix}
\end{align*}
where $f_{\lambda,\rm{model}} = \texttt{norm} \cdot f_{\lambda,\rm{model}}\left(T_{\rm{eff}}, \log{g}, \rm{[M/H]}, \xi_{\rm{mtb}}\right)$ is rescaled to match the flux level of the BH* stack.

Finally, we sample the six parameters $\left( T_{\rm{eff}}, \, \log{g}, \, \text{[M/H]}, \, \xi_{\rm{mtb}}, \, s, \, \texttt{norm} \right)$ with nested sampling via \texttt{dynesty} \citep[multi-ellipsoidal bounds, random-walk sampling;][]{Speagle19}. Parameter posterior medians and 16$^{\rm{th}}$/84$^{\rm{th}}$ percentiles shown on the corner plots (Figs. \ref{fig:atm_fits_stack}, \ref{fig:corner_luminous_substack}, \ref{fig:corner_inter_substack}, \ref{fig:corner_faint_substack}) are importance-weighted quantiles over the full set of nested samples. We then draw $1{,}000$ equal-weight posterior samples by importance resampling, and take the per-wavelength median and quantiles of their model spectra. The effective sample size of the nested sampling weights exceeds $10{,}000$ in every fit, so duplication from resampling is negligible. Fig. \ref{fig:atm_fits_stack} shows the draws for the median BH* stack, and Fig. \ref{fig:atm_fits_substacks} for the substacks.

\subsection{Pseudo-Photospheric Parameters}
\label{sec:parameters}

The pseudo-photospheric parameters are well constrained by the continuum shape of the BH* SEDs, as shown in Fig. \ref{fig:atm_fits_stack} for the median stack. To further demonstrate that the continuum shape actually encodes information about the parameters, we vary a single parameter ($\log{g}$ or [M/H]) while keeping all other parameters fixed at their posterior medians. Fig. \ref{fig:vary_parameter} shows that the model SED is sensitive to variations of these parameters. At lower $\log{g}$ and thus lower pseudo-photospheric density, the near-IR opacity from H$^-$ falls off more steeply than the optical opacity from metal lines. [M/H] impacts the SED through various mechanisms, e.g., by changing metal absorption opacities and by contributing free electrons for H$^-$ formation that dominates the near-IR opacity at these temperatures. Fitting the SED shape can therefore constrain these parameters.

All the parameters of the BH* stacks are summarized in Table \ref{tab:substacks}. In particular, $L_{5500}$ is the criterion used to split the substacks; $T_{\rm{eff}}$ and $\log{g}$ are obtained from the atmosphere model fits; the pseudo-photospheric radius $R_{\rm{phot}}$ is derived using the Stefan-Boltzmann law; $L_{\rm{bol}}$ is obtained from the full-wavelength atmosphere model at each of the $1{,}000$ posterior draws.

The effective temperatures $T_{\rm{eff}}$ obtained from the model fits reside in a much narrower range than those obtained from modified blackbody fits. For the substacks, the model fits yield $T_{\rm{eff}} \approx 4757, \, 4716, \, 4233$ K, spanning only $\approx 500$ K (Fig. \ref{fig:atm_fits_substacks}). In contrast, modified blackbody fits (following \citealt{degraaff25pop}) yield $T_{\rm{eff,\,BB}} \approx 5730, \, 3833, \, 3348$ K, spanning $> 2000$ K. The narrower range of $T_{\rm{eff}}$ reflects the fact that the atmosphere models, like any real atmosphere, are not perfect blackbodies: their continua are affected by both density and metallicity, effects that must be taken into account to measure accurate temperatures. By simultaneously accounting for temperature, density, and metallicity, these model fits are thus more principled than simple/modified blackbodies.

\begin{figure*}[p]
    \centering
    \includegraphics[width=\linewidth]{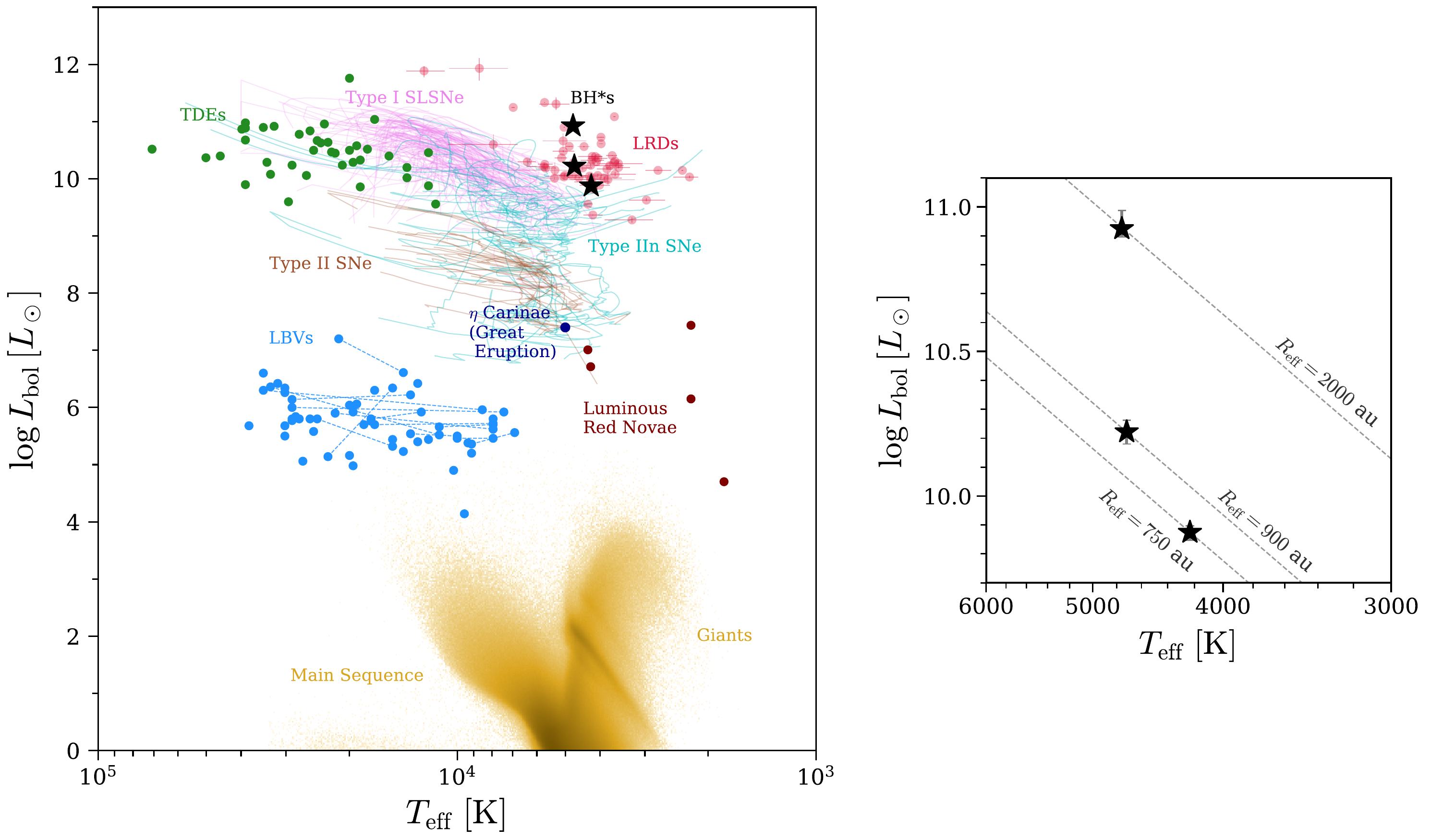}
    \makeatletter
    \def\thempfootnote{\arabic{mpfootnote}}
    \skip\@mpfootins=7pt
    \def\minipagefootnote@here{\par\@ifvoid\@mpfootins{}{\vskip\skip\@mpfootins\hrule height 0.4pt width 0.3\linewidth\kern 4pt\fullinterlineskip\unvbox\@mpfootins}}%
    \makeatother
    \caption{\textbf{LRDs/BH*s occupy the coldest yet most luminous corner (top right) on the Hertzsprung-Russell (HR) diagram of known objects with photospheres or pseudo-photospheres.} BH*s span $\sim 1$ dex in $L_{\rm{bol}}$ while remaining confined to $T_{\rm{eff}} \approx 4200 - 4800$ K. This narrow range of $T_{\rm{eff}}$ is what one would expect if the pseudo-photosphere sits in a dense, optically thick wind. The wind's opacity couples radiation to gas and drives it outward, which requires hydrogen to stay ionized for electron scattering. Expansion cools the wind until hydrogen recombines, at which point the opacity falls sharply and the driving stalls. The pseudo-photosphere is left near that radius, so the narrow temperature range tracks hydrogen recombination \citep[][]{Owocki16}. For luminous blue variables (LBVs), each dashed line connects the minimum- and maximum-light states of a star during its S Doradus phase characterized by slow pulsations.\footnote[3]{Data sources for the HR diagram: TDEs \citep[][]{Gezari21TDE}; Type I superluminous supernovae \citep[SLSNe; from $\sim 50$ rest-frame days before peak to $\sim 250$ days after peak;][]{Gomez24TypeISLSNe}; Type IIn SNe \citep[from $\sim 15-200$ rest-frame days before peak to $\sim 40-500$ days after peak;][]{Hiramatsu24}; Type II SNe \citep[starting from $0-13$ days since explosion;][]{Faran18TypeIISNe}; the Great Eruption of $\eta$ Carinae \citep[][]{Rest12}; LBVs \citep[][]{vanGenderen01SDoradus}; luminous red novae \citep[$200$ days since main peak;][]{Kaminski26rednovae}; LRDs \citep[][]{degraaff25pop}; main sequence and giants \citep[][]{GaiaDR2}.}}
\label{fig:hr_diagram}
\end{figure*}

The $T_{\rm{eff}}$ and $L_{\rm{bol}}$ from the model fits place BH*s in a unique corner on the Hertzsprung-Russell (HR) diagram (Fig. \ref{fig:hr_diagram}). They span $\sim 1$ dex in $L_{\rm{bol}}$ while remaining confined to $T_{\rm{eff}} \approx 4200 - 4800$ K. To contextualize the position of BH*s on the HR diagram, we plot them alongside other known classes of objects with photospheres or pseudo-photospheres. A pseudo-photosphere arises in phenomena like winds and explosions that obscure the central engine. Supernovae (SNe) and luminous red novae have pseudo-photospheres in their expanding ejecta; LBVs in eruption have pseudo-photospheres in an optically thick wind; and tidal disruption events (TDEs) are thought to have reprocessing envelopes around their central engines. Compared to these classes, BH*s are at the coldest yet most luminous corner.

The pseudo-photospheric radii $R_{\rm{phot}}$ of order $\sim 1000$ au appear at first to be in tension with the scales inferred from photoionization modeling, but these are not contradictory. Photoionization models like \texttt{CLOUDY} \citep[e.g.,][]{Ji25BT, Naidu26BHstar, Torralba25IFU} find that the dense gas responsible for electron scattering and Balmer absorption forms a layer $<100$ au thick, the scale implied by their modeled $N_H/n_H$. However, such models effectively treat only a portion of the total structure, i.e., the dense layer where much of the line emission and absorption is imprinted. This dense gas layer is likely a part of a much larger envelope. More detailed radiative-transfer calculations \citep[e.g.,][]{Chang26, Sneppen26, Martins26} will be required to determine whether these regions can be incorporated self-consistently within a single envelope/wind model.

\begin{deluxetable*}{lccccccc}
\tabletypesize{\footnotesize}
\tablecaption{Properties of BH* Stacks}
\tablehead{
\colhead{} & \colhead{$\log\left(L_{5500}/\rm{erg\ s^{-1}}\right)$} & \colhead{$T_{\rm{eff}}/$K} & \colhead{$\log{g}$} & \colhead{$R_{\rm{phot}}/$au} & \colhead{$\log(L_{\rm{bol}}/\rm{erg\ s^{-1}})$} & \colhead{$\log\left(M_\star/M_\odot\right)$} & \colhead{$v_{\rm{blue}, 95\%}/\rm{km\ s}^{-1}$}}
\startdata
\label{tab:substacks}
\vspace{-0.3cm}\\
Median BH* ($N=117$) & $43.5^{+0.1}_{-0.1}$ & $4662^{+27}_{-15}$ & $-2.2^{+0.2}_{-0.2}$ & $941^{+64}_{-94}$ & $43.8^{+0.1}_{-0.1}$ & $8.5^{+0.1}_{-0.2}$ & $-495^{+95}_{-92}$ \\
\noalign{\vskip 2pt}
\hline
\noalign{\vskip 2pt}
Luminous BH* ($N=17$) & $44.2^{+0.1}_{-0.1}$ & $4757^{+37}_{-20}$ & $-2.5^{+0.1}_{-0.1}$ & $1989^{+149}_{-77}$ & $44.5^{+0.1}_{-0.1}$ & $8.2^{+0.2}_{-0.2}$ & $-523^{+121}_{-251}$ \\
Inter. BH* ($N=85$) & $43.5^{+0.1}_{-0.1}$ & $4716^{+38}_{-40}$ & $-1.9^{+0.9}_{-0.4}$ & $900^{+50}_{-43}$ & $43.8^{+0.1}_{-0.1}$ & $8.0^{+0.1}_{-0.1}$ & $-364^{+185}_{-195}$ \\
Faint BH* ($N=15$) & $42.9^{+0.1}_{-0.1}$ & $4233^{+23}_{-21}$ & $-2.5^{+0.3}_{-0.3}$ & $747^{+23}_{-26}$ & $43.5^{+0.1}_{-0.1}$ & $8.4^{+0.1}_{-0.1}$ & $-$
\enddata
\tablecomments{$L_{5500}$ is the criterion used to split the substacks. $T_{\rm{eff}}$ and $\log{g}$ are obtained from the atmosphere model fits. $R_{\rm{phot}}$ is derived using the Stefan-Boltzmann law. $L_{\rm{bol}}$ is obtained from the full-wavelength atmosphere model at each of the $1{,}000$ posterior draws. Stellar mass $M_\star$ of the host galaxy is obtained from \texttt{Prospector} modeling (see Fig. \ref{fig:host_substacks_prospector} in the Appendix). $v_{\rm{blue}, 95\%}$ traces the blue edge of the absorber in LRDs -- the velocity at which transmission recovers to $95\%$. Of the $18$ absorbers compiled in \cite{Matthee26}, $11$ have $v_{\rm{blue}, 95\%}$ reported in \cite{Naidu26}. Eight fall in the luminous substack, two in the intermediate, and none in the faint. We perturb each $v_{\rm{blue}, 95\%}$ measurement and bootstrap resample over $1{,}000$ trials to obtain the uncertainty on the sample median.}
\end{deluxetable*}

\section{Results: BH* Mass Estimates}
\label{sec:results}

Following \cite{Naidu26}, we estimate the BH* mass using four different methods, which we describe below. The derived masses for the median BH* stack as well as the substacks are summarized in Tables \ref{tab:mass_upper_bounds}, \ref{tab:mass}, and Figs. \ref{fig:virial_comparison}, \ref{fig:m_vs_mstar}. We emphasize that the mass estimates here should be treated as exploratory order-of-magnitude estimates. Standard virial estimators may be inapplicable to LRDs, and we are in need of fresh estimates that go a step beyond pointing out that the standard estimators are inapplicable. Given this landscape, we find even order-of-magnitude mass estimates to be insightful.

\subsection{Surface Gravity}
\label{sec:surface_gravity}

The atmosphere model fitting yields self-consistent $\log{g}$, $T_{\rm{eff}}$, and $L_{\rm{bol}}$. Together with the pseudo-photospheric radius $R_{\rm{phot}}$ derived from the Stefan-Boltzmann law, these parameters give the BH* mass,
\begin{equation}
    M_{\rm{BH^*}} \, = \, \frac{g R_{\rm{phot}}^2}{G} \, \begin{matrix} \\.\\ \end{matrix}
\label{eq:surface_gravity}
\end{equation}

As discussed in \S\ref{sec:models}, the models constrain the net gravity $g - g_{\rm{dyn}}$, where $g_{\rm{dyn}}$ accounts for dynamical support due to gas motion \citep[][]{Liu26TLUSTY}. LRDs often show deep P-Cygni profiles in \ion{He}{1} and Balmer lines indicative of a blueshifted outflow \citep[e.g.,][]{Wang25Outflow, Loiacono25, Kokorev25glimpsed, Matthee26}. Further, the Balmer series blueshifts in the median LRD decrease in magnitude as one goes from H$\alpha$ to H$\beta$ \citep[e.g.,][de Graaff et al. in prep., Katz et al. in prep.]{Matthee26}, also characteristic of an outflowing medium where transitions emerging farther out are more blueshifted. If we are observing an escaping outflow, we expect $g_{\rm{dyn}} < 0$, so our fitted $g$ would exceed the true surface gravity and the mass inferred from it would be correspondingly overestimated. On the other hand, if instead we are observing e.g. an equatorial accretion flow \citep[e.g.,][]{Matthee26}, we expect $g_{\rm{dyn}} > 0$. There is evidence for both regimes in LRDs -- the H$\alpha$ absorption in $\approx 80-90\%$ of LRDs is blueshifted, and in $\approx 10-20\%$ is redshifted \citep[e.g.,][]{Yanagisawa26c, Lin26, Matthee26, Juodzbalis26, Davis26}. The redshifts are generally of lower magnitude than the blueshifts. For our fiducial mass estimates, we assume $g_{\rm{dyn}} \approx 0$ for simplicity, since we are dealing with stacked spectra covering both regimes. In detail, when the medium is net outflowing, the mass estimate is an upper limit; when the medium is net inflowing, the mass estimate is a lower limit.

In Fig. \ref{fig:vary_parameter}, we demonstrate that the spectral shape contains information about $\log{g}$. We find $\log{g}$ ranging from $-2.5$ to $-1.9$. This is two orders of magnitude lower than the minima seen in ``true'' stars such as hypergiants \citep[e.g.,][]{Lin26}. Such low surface gravities are consistent with the vast radial span expected for BH* pseudo-photospheres \citep[e.g.,][]{Liu25BB, Kido25, Nandal26}. Based on these values, for the median BH*, we obtain $\log{\left(M_{\rm{BH^*}}/M_\odot\right)} \approx 4.0$.

\begin{deluxetable*}{lcccc}
\tabletypesize{\footnotesize}
\tablecaption{BH* Mass $\log{M_{\rm{BH^*}}}$ Conservative Upper Bounds}
\tablehead{
\colhead{} & \colhead{Surface gravity (\S\ref{sec:surface_gravity})} & \colhead{$L_{\rm{bol}}/L_{\rm{Edd}} > 1$ (\S\ref{sec:eddington_lum})} & \colhead{Escape velocity (\S\ref{sec:escape_velocity})} & \colhead{Variability (\S\ref{sec:variability})}}
\startdata
\label{tab:mass_upper_bounds}
\vspace{-0.3cm} \\
Median BH* ($N=117$) & $-$ & $5.7^{+0.1}_{-0.1}$ & $5.1^{+0.2}_{-0.2}$ & $5.0^{+0.1}_{-0.2}$ \\
\noalign{\vskip 2pt}
\hline
\noalign{\vskip 2pt}
Luminous BH* ($N=17$) & $-$ & $6.4^{+0.1}_{-0.1}$ & $5.5^{+0.4}_{-0.3}$ & $6.0^{+0.1}_{-0.1}$ \\
Inter. BH* ($N=85$) & $-$ & $5.7^{+0.1}_{-0.1}$ & $4.8^{+0.4}_{-0.7}$ & $5.0^{+0.1}_{-0.1}$ \\
Faint BH* ($N=15$) & $-$ & $5.4^{+0.1}_{-0.1}$ & $-$ & $4.7^{+0.1}_{-0.1}$
\enddata
\tablecomments{BH* mass is bounded using four approaches, reported in units of $M_{\odot}$. See text for details on the various arguments. The errors are statistical and do not capture systematics.}
\end{deluxetable*}

\begin{deluxetable*}{lcccc}
\tabletypesize{\footnotesize}
\tablecaption{BH* Mass $\log{M_{\rm{BH^*}}}$ Fiducial Estimates}
\tablehead{
\colhead{} & \colhead{Surface gravity (\S\ref{sec:surface_gravity})} & \colhead{$L_{\rm{bol}}/L_{\rm{Edd}} = 5-50$ (\S\ref{sec:eddington_lum})} & \colhead{Escape velocity (\S\ref{sec:escape_velocity})} & \colhead{Variability (\S\ref{sec:variability})}}
\startdata
\label{tab:mass}
\vspace{-0.3cm} \\
Median BH* ($N=117$) & $4.0^{+0.2}_{-0.2}$ & $4.0^{+0.1}_{-0.1} - 5.0^{+0.1}_{-0.1}$ & $4.2^{+0.2}_{-0.2}$ & $4.1^{+0.1}_{-0.2}$ \\
\noalign{\vskip 2pt}
\hline
\noalign{\vskip 2pt}
Luminous BH* ($N=17$) & $4.3^{+0.1}_{-0.2}$ & $4.7^{+0.1}_{-0.1} - 5.7^{+0.1}_{-0.1}$ & $4.5^{+0.4}_{-0.3}$ & $5.0^{+0.1}_{-0.1}$ \\
Inter. BH* ($N=85$) & $4.2^{+0.9}_{-0.4}$ & $4.0^{+0.1}_{-0.1} - 5.0^{+0.1}_{-0.1}$ & $3.9^{+0.4}_{-0.7}$ & $4.0^{+0.1}_{-0.1}$ \\
Faint BH* ($N=15$) & $3.4^{+0.4}_{-0.3}$ & $3.7^{+0.1}_{-0.1} - 4.7^{+0.1}_{-0.1}$ & $-$ & $3.8^{+0.1}_{-0.1}$
\enddata
\tablecomments{BH* mass is reported in units of $M_{\odot}$. The errors are statistical and do not capture systematics.}
\end{deluxetable*}

\subsection{$L_{\rm{bol}}/L_{\rm{Edd}}$}
\label{sec:eddington_lum}

Various models that attempt to reproduce LRD features require close-to-Eddington or super-Eddington accretion that is often accompanied by super-Eddington luminosities \citep[e.g.,][]{Pacucci24, Inayoshi25, Kido25, Liu25BB, Madau26, Sneppen26, Inayoshi26specuni, Lambrides26xrays}. The massive star phenomena whose spectral features resemble those of LRDs are characterized by highly super-Eddington luminosities -- e.g., the Great Eruption of $\eta$ Carinae shone at $\Gamma_{\rm{es}} \approx 5$ (where the Eddington parameter $\Gamma_{\rm{es}} = L_{\rm{bol}}/L_{\rm{Edd}}$) for two decades \citep[][]{Smith18}, while Type IIn SNe and luminous red novae can reach $\Gamma_{\rm{es}} \approx 50$ \citep[e.g.,][]{Davidson16, Smith16, Cai22}. This motivates estimating the mass of the BH* from its Eddington luminosity,
\begin{equation}
    M_{\rm{BH^*}} \, = \, \frac{\kappa_{\rm{es}} L_{\rm{Edd}}}{4\pi Gc} \, = \, \frac{\kappa_{\rm{es}}}{4\pi Gc} \cdot \frac{L_{\rm{bol}}}{\Gamma_{\rm{es}}} \begin{matrix} \\,\\ \end{matrix}
\label{eq:eddington_lum}
\end{equation}
where $\kappa_{\rm{es}} = 0.40 \, \rm{cm}^2 \, \rm{g}^{-1}$ is the electron-scattering opacity of fully ionized hydrogen. Note that we assume electron scattering is the dominant source of opacity for this calculation. This is consistent with the values $\Gamma_{\rm{es}} \approx 5 - 50$ quoted above for $\eta$ Carinae, Type IIn SNe and luminous red novae, which are derived under the same assumption. Similar approaches have been used to estimate LRD central engine masses in previous work \citep[e.g.,][]{Greene25, Umeda25, Gentile26, Naidu26, Yanagisawa26}.

Adopting the conservative bound $\Gamma_{\rm{es}} > 1$ yields $\log{\left(M_{\rm{BH^*}}/M_\odot\right)} < 5.7$ for the median BH*. For our fiducial estimate, we adopt the range $\Gamma_{\rm{es}} \approx 5 - 50$ by analogy with massive star eruptions and obtain $\log{\left(M_{\rm{BH^*}}/M_\odot\right)} \approx 4.0 - 5.0$.

\subsection{Escape Velocity}
\label{sec:escape_velocity}

P-Cygni signatures in the \ion{He}{1} and Balmer line profiles of the typical LRD indicate a net outflowing medium \citep[e.g.,][]{NM24, Juodzbalis24rosetta, Wang25Outflow, Matthee26}. We define the local escape velocity at every radius, $v_{\rm{esc}}(r) = \sqrt{2GM_{\rm{BH^*}}/r}$. Consider the local escape velocity at $R_{\rm{phot}}$,
$v_{\rm{esc}}\left(R_{\rm{phot}}\right)$. In an optically thick massive-star wind, as the outflow is accelerated deep in the interior, it reaches its terminal velocity $v_\infty$ well below the pseudo-photosphere \citep[e.g.,][]{Schillemans26}. At and beyond the pseudo-photosphere, the temperature has dropped sufficiently for hydrogen to recombine and for opacity to decrease, so further acceleration is minimal, i.e., $v_\infty > v_{\rm{esc}}\left(R_{\rm{phot}}\right)$. Note that this assumes the wind experiences no deceleration -- e.g., an entirely ``failed wind'' that decelerates along all sightlines and falls back would not achieve this inequality. 

We do not have a direct measurement of $v_\infty$, so we approximate it with $v_{\rm{blue}, 95\%}$ at the blue edge of the absorption trough. This is defined as the velocity at which the transmission recovers to $95\%$ \citep[][]{Torralba26panbhstar, Naidu26}. Rearranging the inequality then gives \citep[][]{Naidu26}:
\begin{equation}
    M_{\rm{BH^*}} \, = \, \frac{R_{\rm{phot}} \, v_{\rm{esc}}^2}{2G} \, < \, \frac{R_{\rm{phot}} \, v_{\rm{blue}, 95\%}^2}{2G} \begin{matrix} \\.\\ \end{matrix}
\label{eq:escape_velocity}
\end{equation}

To estimate $v_{\rm{blue}, 95\%}$, we use the set of $18$ absorbers compiled in \cite{Matthee26}, of which $11$ have $v_{\rm{blue}, 95\%}$ reported in \cite{Naidu26}. Of these, eight fall in the luminous substack with $v_{\rm{blue}, 95\%} \approx -523^{+121}_{-251} \, \rm{km} \, \rm{s}^{-1}$, two in the intermediate with $v_{\rm{blue}, 95\%} \approx -364^{+185}_{-195} \, \rm{km} \, \rm{s}^{-1}$, and none in the faint (so no estimate is quoted for the faint substack in Tables \ref{tab:mass_upper_bounds}, \ref{tab:mass}). We bootstrap resample $v_{\rm{blue}, 95\%}$ to obtain the uncertainties on the medians. Note that even in cases where the absorption centroid is redshifted, the trough can still extend to the blueshifted region \citep[e.g.,][]{Naidu26}, so $v_{\rm{blue}, 95\%}$ remains applicable.

For the median BH*, we find the upper bound $\log{\left(M_{\rm{BH^*}}/M_\odot\right)} < 5.1$. In massive-star winds, we typically have $v_\infty / v_{\rm{esc}} \approx 3$ \citep[e.g.,][]{Lamers95, MullerVink08, Hawcroft24}, which would overestimate the mass by a factor of $\approx 9 \times$. Indeed, \cite{Naidu26} found that this approach overestimates the mass of $\eta$ Carinae by a factor of $\approx 10 \times$. Adopting the $\approx 9 \times$ correction, we arrive at our fiducial $\log{\left(M_{\rm{BH^*}}/M_\odot\right)} \approx 4.2$.

\subsection{Variability}
\label{sec:variability}

We take the dynamical time of the extended pseudo-photosphere to be $t_{\rm{dyn}} = R_{\rm{phot}} / v_{\rm{esc}}$, the timescale on which the structure can respond to a perturbation. Rewriting $v_{\rm{esc}}$ in terms of $R_{\rm{phot}}$ and $M_{\rm{BH^*}}$ gives $t_{\rm{dyn}}^2 = R_{\rm{phot}}^3 / 2GM_{\rm{BH^*}}$, so we have
\begin{equation}
    M_{\rm{BH^*}} \, = \, \frac{R_{\rm{phot}}^3}{2G t_{\rm{dyn}}^2} \begin{matrix} \\.\\ \end{matrix}
\label{eq:variability}
\end{equation}

The typical LRD shows little to no optical variability over $\gtrsim 10$ yr baselines \citep[e.g.,][]{Kokubo25, Tee25, Zhang25var, Liu26TWINKLE, Burke26, Stone26}. This has been established on a population level through repeat measurements that crucially use the same instrumental mode, so that any apparent variability is not the product of e.g. slit-losses or cross-instrument calibration uncertainties \citep[e.g.,][]{Tee25, Liu26TWINKLE, Stone26, Park26}. While some studies report tentative variability on $< 10$ rest-frame yr timescales in a handful of LRDs often using slit-based spectroscopy \citep[e.g.,][]{Ji25BT, Furtak25, Naidu26BHstar, Lambrides26, Brooks26, DEugenio26z5lrd}, it remains to be seen whether these signals survive after accounting for position-dependent slit variations, which are hard to quantify for LRDs that show strong radial color gradients. Given the sharp difference between the point source and extended source \citep[e.g.,][]{Ishikawa26}, each subtly different slit position might produce different spectra mimicking variability.

Adopting $t_{\rm{dyn}} \gtrsim 10$ yr as a conservative bound yields $\log{\left(M_{\rm{BH^*}}/M_\odot\right)} < 5.0$ for the median BH*. A sharper estimate follows from the lensed LRD R2211-RX1, whose multi-epoch photometry is consistent with a $t_{\rm{dyn}} \approx 30$ yr envelope pulsation \citep[][]{Zhang25cepheid, Zhang26H0, Cantiello25}. This gives our fiducial estimate $\log{\left(M_{\rm{BH^*}}/M_\odot\right)} \approx 4.1$.

\subsection{Summary of Mass Estimates}
\label{sec:mass_summary}

\begin{figure}
    \centering
    \includegraphics[width=0.8\linewidth]{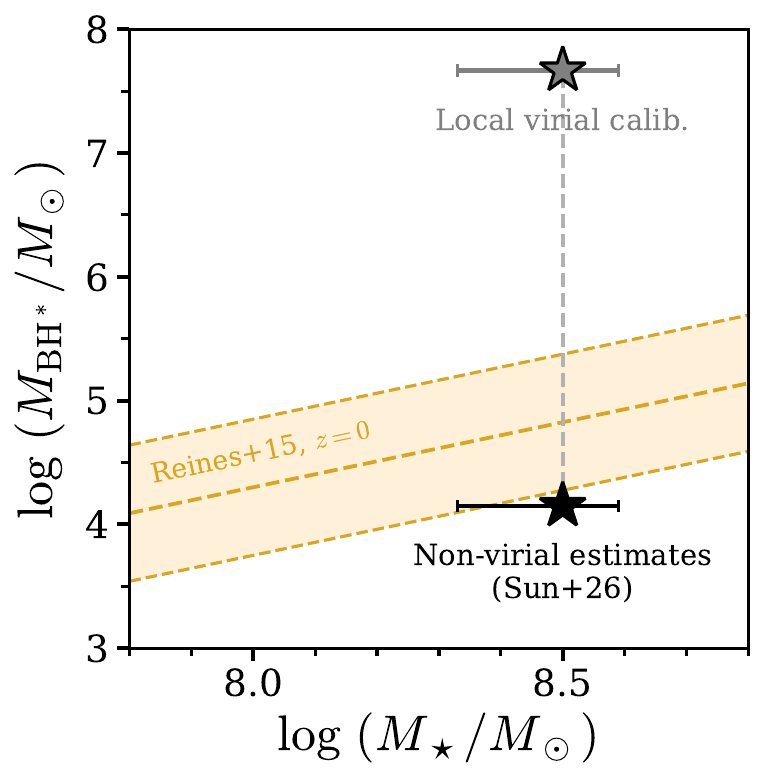}
    \caption{\textbf{The non-virial BH* mass is $\gtrsim 3$ dex lower than that derived from the local virial calibrations \citep{Reines13} -- BH* mass vs. host stellar mass for the median stack.} The bottom datapoint shows the median of the non-virial masses derived from the four approaches. We only show the median mass here for visual clarity; Table \ref{tab:mass} shows that the four approaches all yield BH* masses $\sim 3$ dex lower than that derived from the virial calibrations.}
\label{fig:virial_comparison}
\end{figure}

All four methods yield consistent fiducial masses of $10^{4-5} \, M_\odot$ for the median BH*, and $\approx 10^{3.5-6} \, M_\odot$ for all the BH* substacks, as summarized in Table \ref{tab:mass}. As per these non-virial estimators, Fig. \ref{fig:virial_comparison} shows that the median BH* has a non-virial mass $\gtrsim 3$ dex lower than that derived from the local virial calibrations \citep[][]{Reines13}.

In Fig. \ref{fig:m_vs_mstar}, we compare the inferred host galaxy stellar mass for each substack from SED fitting against the BH* mass. The stellar mass $M_\star$ of the host galaxy is obtained from \texttt{Prospector} modeling \citep[][]{Leja17, Leja19, Johnson21} following the choices described in \citet[][]{Sun26}, in turn adapted from \cite{Tacchella22}, \cite{NM24}. The SED fits and the accompanying star formation histories are included in the Appendix (Fig. \ref{fig:host_substacks_prospector}). The host stellar masses are $\approx 10^8 \, M_\odot$, consistent with independent clustering-based arguments \citep[e.g.,][]{Matthee25LRDclustering, Lin25clustering}.

Figs. \ref{fig:virial_comparison} and \ref{fig:m_vs_mstar} show that, under these estimators, the typical BH*s do not outweigh their host galaxies and are not ``overmassive.'' Instead, most of our estimates are at or below the $z=0$ $M_{\rm{BH}}$ vs. $M_\star$ relation. Among the BH*s, the more luminous BH*s are more massive. The host stellar masses, however, do not follow the same trend, so there is no monotonic ordering between the stellar mass, BH* luminosity, and BH* mass. One interpretation is that we are witnessing the very early stages of the assembly of this relation, where more or less massive BH*s can form within more or less massive galaxies. Further, BH*s could grow significantly from high redshift to $z=0$ via BH-BH mergers \citep[e.g.,][]{Tanaka24, Merida25, Yanagisawa26} or secular growth (e.g., galaxy mergers, accretion), so they might appear ``undermassive,'' sitting below the $z=0$ $M_{\rm{BH}}$ vs. $M_\star$ relation.

\subsection{Systematic Uncertainties}
\label{sec:systematic_uncertainties}

\begin{figure*}
    \centering
    \includegraphics[width=0.70\linewidth]{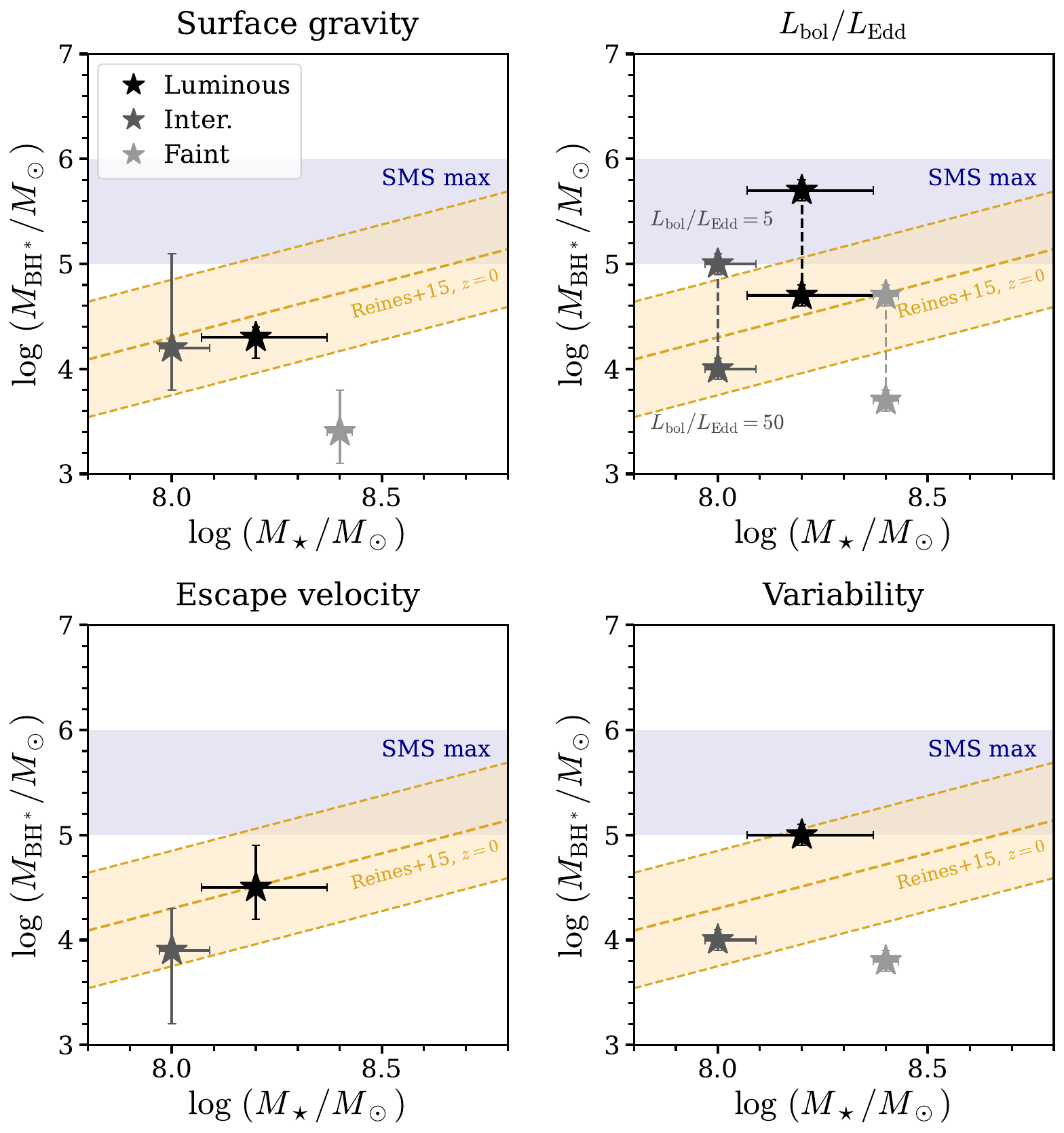}
    \caption{\textbf{Non-virial BH* masses are consistent with BH*s arising from single supermassive stars (SMSs) -- BH* mass vs. host stellar mass for the three substacks.} Each panel shows the fiducial masses (Table \ref{tab:mass}) derived from one of the four approaches described in \S\ref{sec:results}, plotted against host stellar masses derived from \texttt{Prospector} fits (Fig. \ref{fig:host_substacks_prospector}). The BH* masses all sit below the maximum theoretical mass of an SMS \citep[blue band, $\approx 10^{5-6} \, M_\odot$ due to general relativistic instabilities; e.g.,][]{Chandrasekhar64, Fowler66, Woods17, Nandal24gr, Saio24gr}, supporting the picture that BH*s may arise from single SMSs. Note that in our proposed picture, the BH* masses may lie below the $z=0$ $M_{\rm{BH}}$ vs. $M_\star$ relation \citep[][]{RV15}, which is plausible as the black hole could grow significantly from high redshift to $z=0$ via BH-BH mergers \citep[e.g.,][]{Tanaka24, Merida25, Yanagisawa26} or secular growth such as accretion.}
\label{fig:m_vs_mstar}
\end{figure*}

We emphasize that the derived masses should be treated as exploratory order-of-magnitude estimates. The uncertainties quoted (Tables \ref{tab:mass_upper_bounds}, \ref{tab:mass}, Fig. \ref{fig:m_vs_mstar}) are statistical and do not capture systematic effects. One such systematic is the LRD $-$ host $=$ BH* decomposition, which may not yet be perfect. The non-monotonic ordering of stellar masses across the substacks (Fig. \ref{fig:m_vs_mstar}) may reflect this, so the host stellar masses reported should be read collectively as spanning a rough range, rather than individually as precise values \citep[e.g.,][]{Leja19, Pacifici23, Wang24}. Still, it is a reassuring check that the derived stellar masses are consistent with clustering studies, which show that LRDs reside in environments typical of low-mass dwarf galaxies of masses $\approx 10^{7-8} \, M_\odot$ \citep[e.g.,][]{Matthee25LRDclustering, Pizzati25, Lin25clustering}.

Another systematic concerns $R_{\rm{phot}}$, as three of our four methods depend on $R_{\rm{phot}}$ and a shift in $R_{\rm{phot}}$ would shift the mass estimates together. One of the ways to probe this is to ask how much $R_{\rm{phot}}$ would change if a small amount of dust were included in the fitting. Observed LRDs show little far-IR dust reemission \citep[e.g.,][]{Casey25, Setton25, Xiao25}, but some low-$z$ LRDs show clear mid-IR excesses relative to a pure pseudo-photosphere \citep[e.g.,][]{Ji25lol, Lin25, Park26} that may be due to free-free emission or a modest amount of dust \citep[e.g.,][]{Naidu26, Ashall26, Kiyota26}. Whether these low-$z$ LRDs are representative of the LRD population remains to be seen, but their mid-IR excesses suggest that a small amount of dust may be present. We therefore repeat the fitting with dust attenuation included, using the Calzetti dust law \citep[][]{Calzetti00}, with V-band attenuation $A_V$ drawn from a uniform prior over $[0, 1]$ mag. The $T_{\rm{eff}}$ spread increases from $\approx 500$ K to $\approx 1000$ K, still much narrower than the $> 2000$ K spread suggested by modified blackbody fits. $R_{\rm{phot}}$ decreases for all BH* substacks, with the largest change for the luminous BH*, where it decreases by $\approx 25\%$. The mass changes by $< 0.5$ dex in almost all cases, except for the luminous BH* where the surface gravity approach gives an increase of $\approx 1.3$ dex -- the increase in mass results from the altered shape of the continuum used to constrain $\log{g}$.

Several paths forward could help better constrain the systematic uncertainties. On the observational side, higher-resolution spectroscopy could improve measurements of the offsets and widths of absorption features \citep[e.g.,][]{DEugenio26z5lrd, Davis26, Matthee26}, which would help finely map the velocity structure of the envelope and e.g., constrain how much of the fitted surface gravity is dynamical versus gravitational. Comparing the blue edges of absorption troughs \citep[e.g.,][]{Torralba26panbhstar} across species would help further test whether the wind has reached its terminal velocity where we observe it. Separately, finding and monitoring more lensed LRDs would expand the sample beyond the single LRD that we rely on now for the dynamical timescale estimate: magnification makes small variations measurable, and time delays in multiply-imaged systems extend the baseline over which the timescale can be measured \citep[e.g.,][]{Zhang25cepheid, Furtak25, Golubchik25}. On the modeling side, self-consistent models for NLTE effects, the wind structure, and the inhomogeneous dense gas \citep[e.g.,][]{Martins26, Sneppen26, Chang26} could predict the pseudo-photospheric parameters and line profiles from one combined physical structure, rather than fitting the continuum and the line-forming layers separately.

\section{Discussion}
\label{sec:discussion}

The mass of the LRD central engine remains debated by several orders of magnitude, spanning from seed black holes to supermassive black holes. While virial estimators place BH*s at $\approx10^{6-8} M_\odot$, our four methods, under the pseudo-photosphere picture, instead converge on $\approx 10^{3.5-6} \, M_\odot$. Given how common LRDs are, the two answers imply very different pictures of how the earliest black holes formed. Now we discuss what follows provided these low BH* masses are correct.

\subsection{Could BH*s Arise from Single Supermassive Stars?}
\label{sec:mass_discussion}

\subsubsection{Motivation}

The similarities between LRDs and massive star eruptions invite us to consider an exciting possibility: that BH*s may form from individual supermassive stars (SMSs; $10^{3-5} \, M_\odot$) observed in a state similar to $\eta$ Carinae during its Great Eruption \citep[e.g.,][]{Nandal26smslrds, Zwick26, Chisholm26, Martins26, Naidu26}. Prolific mass loss of optically thick enshrouding material is a ubiquitous phenomenon among massive stars \citep[e.g.,][]{SmithOwocki06, Smith14, Morris17, Sabhahit26} -- SMSs, if they exist, would likely be no different as per theoretical models \citep[e.g.,][]{Nandal26}. Such mass loss would elegantly produce a fully covering shroud of dense gas ($f_{\rm{cov}}\approx100\%$; e.g., \citealt[][]{Yan25, Yanagisawa4pi, Kageura26}). This would build a BH* with close to unity covering fractions ``inside out'' instead of having to deposit gas with a high $f_{\rm{cov}}$ on an existing black hole ``outside in'' \citep[e.g.,][]{Alexander14}.

Material shed by an SMS is predicted to carry the imprint of CNO-processed hydrogen burning and thus envelop the star in nitrogen-rich gas \citep[e.g.,][]{Denissenkov14, Charbonnel23, Nagele23, Marques-Chaves24, Nandal25Nitrogen}. Intriguingly, LRDs do show highly elevated [N/C] \citep[e.g.,][]{Isobe25, Morel25, Ji26nitrogen, Papovich26}. This unique abundance pattern is thought to be a generic feature of star formation in dense, massive clusters \citep[e.g.,][]{Belokurov23, Belokurov24, Naidu25z14, Schaerer25GCs, Ji26nitrogen}. In such clusters, rapid runaway collisions may be rampant, resulting in very massive stars (VMSs) and SMSs \citep[e.g.,][]{Boekholt18, Gieles18, Rantala26, Williams26, Chung26, Tanikawa26}. Indeed, VMSs $\gtrsim500 M_{\rm{\odot}}$ appear to be relatively common in the early Universe \citep[e.g.,][]{Marques-Chaves26, Yang26VMS, Chen26VMSGNz11}. Might BH*s mark the SMSs?

\subsubsection{Non-Virial Mass Estimates Allow for SMS Progenitors}

Our derived BH* masses overlap with those of SMSs in theoretical models that propose SMSs as heavy black hole seeds. In these calculations, a star accreting at $\gtrsim 0.1 \, M_\odot \, \rm{yr}^{-1}$ grows to $10^{4-5} \, M_\odot$ before the general relativistic instability triggers its collapse, leaving a black hole of comparable mass \citep[e.g.,][]{Hosokawa13, Umeda16, Nandal26SMSlifetimes}. While virial mass estimates ($\approx 10^{7-8} \, M_\odot$) rule out SMSs or their immediate descendants, our estimates fall well within this range, i.e., the mass a heavy seed would have at birth. Crucially, the instability sets an SMS mass ceiling of $\approx 10^{5-6} \, M_\odot$ \citep[][]{Chandrasekhar64, Fowler66, Woods17, Nandal24gr, Saio24gr}, and all of our estimates in Fig. \ref{fig:m_vs_mstar} fall below it -- a necessary condition for BH*s to arise from SMSs. Given the numerous parallels between BH*s and massive star ($\approx100 \, M_\odot$) phenomena noted in \S\ref{sec:intro}, it is a natural hypothesis that BH*s may be a scaled-up SMS version of those phenomena.

\subsubsection{The Maximum SMS Mass as the Bright-End Cutoff of the LRD Luminosity Function}

Reconciling the low BH* masses with the derived bolometric luminosities requires strongly super-Eddington luminosity. Since $M_{\rm{BH^*}} = \kappa_{\rm{es}} L_{\rm{bol}} / 4\pi Gc \Gamma_{\rm{es}}$ (Eq. \ref{eq:eddington_lum}), inserting the derived $L_{\rm{bol}}$ together with the masses from the three non-$L_{\rm{bol}}/L_{\rm{Edd}}$ methods yields $\Gamma_{\rm{es}} = L_{\rm{bol}}/L_{\rm{Edd}} \sim 50$. At this high Eddington ratio, a maximal $\sim 10^{5-6} \, M_\odot$ SMS would reach $L_{\rm{bol}} \sim 10^{44.8-45.8} \, \rm{erg\ s}^{-1}$, which coincides with the sharp bright-end cutoff of the LRD luminosity function at $\approx10^{45-46} \, \rm{erg\ s}^{-1}$ \citep[e.g.,][]{Ma25counting, Weibel26}. So the SMS mass ceiling may offer an elegant explanation for the cutoff of the observed LRD luminosities. 

While the derived $L_{\rm{bol}}/L_{\rm{Edd}}$ may seem high at first glance, this is common among enshrouded eruptions. Further, note that we are reporting luminosities in excess of the Eddington luminosity, but not necessarily super-Eddington accretion rates. In the SMS picture, the dense gas cocoons enable enormous luminosities through highly efficient conversion of kinetic energy into radiative luminosity ($>50\times$ more efficient than ordinary SNe; e.g., \citealt{Smith17}) via circumstellar medium interaction as proposed in \citet[][]{Naidu26}. Circumstellar medium interaction accounts for some of the brightest classes of transients (e.g., Type IIn SNe; see Fig. \ref{fig:hr_diagram}) and may also help explain the dramatic luminosities of LRDs despite their low-mass central engines.

\subsubsection{The Fundamental ``Stellar'' Parameters of BH*s and SFHs of LRD Hosts are Consistent with SMS}

Indeed, our derived parameters are consistent with a physical picture in which BH*s originate from SMSs with winds. Pulsational ejection from an SMS envelope can assemble a compact, optically thick shell \citep[][]{Nandal26} with LRD-like chemical abundances \citep[e.g.,][]{Isobe25, Morel25, Ji26nitrogen, Papovich26}. Super-Eddington radiation drives the wind whose opacity couples the radiation to the gas and pushes it outward. This coupling persists only while hydrogen stays ionized and free electrons can scatter. As the wind expands, it cools until hydrogen recombines, at which point the opacity falls sharply and the driving stalls, leaving the pseudo-photosphere near that radius. Therefore, $T_{\rm{eff}}$ tracks hydrogen recombination and is expected to fall within a narrow range of $5000-6000$ K \citep[][]{Owocki16, Nandal26}. This is close to what we find: in the HR diagram (Fig. \ref{fig:hr_diagram}), BH*s remain confined to $T_{\rm{eff}} \approx 4200-4800$ K.

The environments in which BH*s reside are consistent with those expected for SMS formation. The low metallicities recovered from the atmosphere model fits, [M/H] $\approx -1.9 - -1.8$, indicate metal-poor environments where SMSs are expected to form \citep[e.g.,][]{Chon26seeds}. We caution that these values sit at the lower bound of the model grid, so they are best taken as evidence for a metal-poor environment rather than as precise measurements. \cite{Chisholm26} show that the redshift evolution of LRD number density -- under plausible conditions -- tracks the evolution of metal-poor ([Fe/H] $< -1.4$) globular clusters in the Milky Way. Globular clusters are where runaway stellar mergers may build SMSs \citep[e.g.,][]{Denissenkov14, Gieles18, Gieles25, Vergara25, Shi26}. Our inferred low metallicities and recent bursty star formation histories for BH*s (\citealt[][]{Sun26}, see also the Appendix for substack SFHs) are thus consistent with the environments expected for SMS formation. 

Runaway collisions need not be the only unique route to SMS formation. The blue companions reported around LRDs \citep[e.g.,][]{Baggen26, Baggen25MI, Pacucci26lrd_companions, Barger26} have been interpreted as Lyman-Werner sources that suppress H$_2$ cooling and precipitate direct collapse of primordial gas into an SMS \citep[e.g.,][]{Naidu26BHstar, Baggen26, Pacucci26lrd_companions, Chon26seeds}. Alternatively, the blue companions may indicate that LRDs form in merger-driven starbursts \citep[e.g.,][]{Sun26, Barger26}. In the SMS picture, the initial rapid assembly of the star is driven by e.g. runaway collisions or direct collapse, and the later growth of the central BH can occur at super-Eddington rates as described in quasistar models \citep[e.g.,][]{Dotan11, Begelman25, Santarelli25, Roman-Garza26, Hassan26}.

\begin{figure*}
    \centering
    \includegraphics[width=\linewidth]{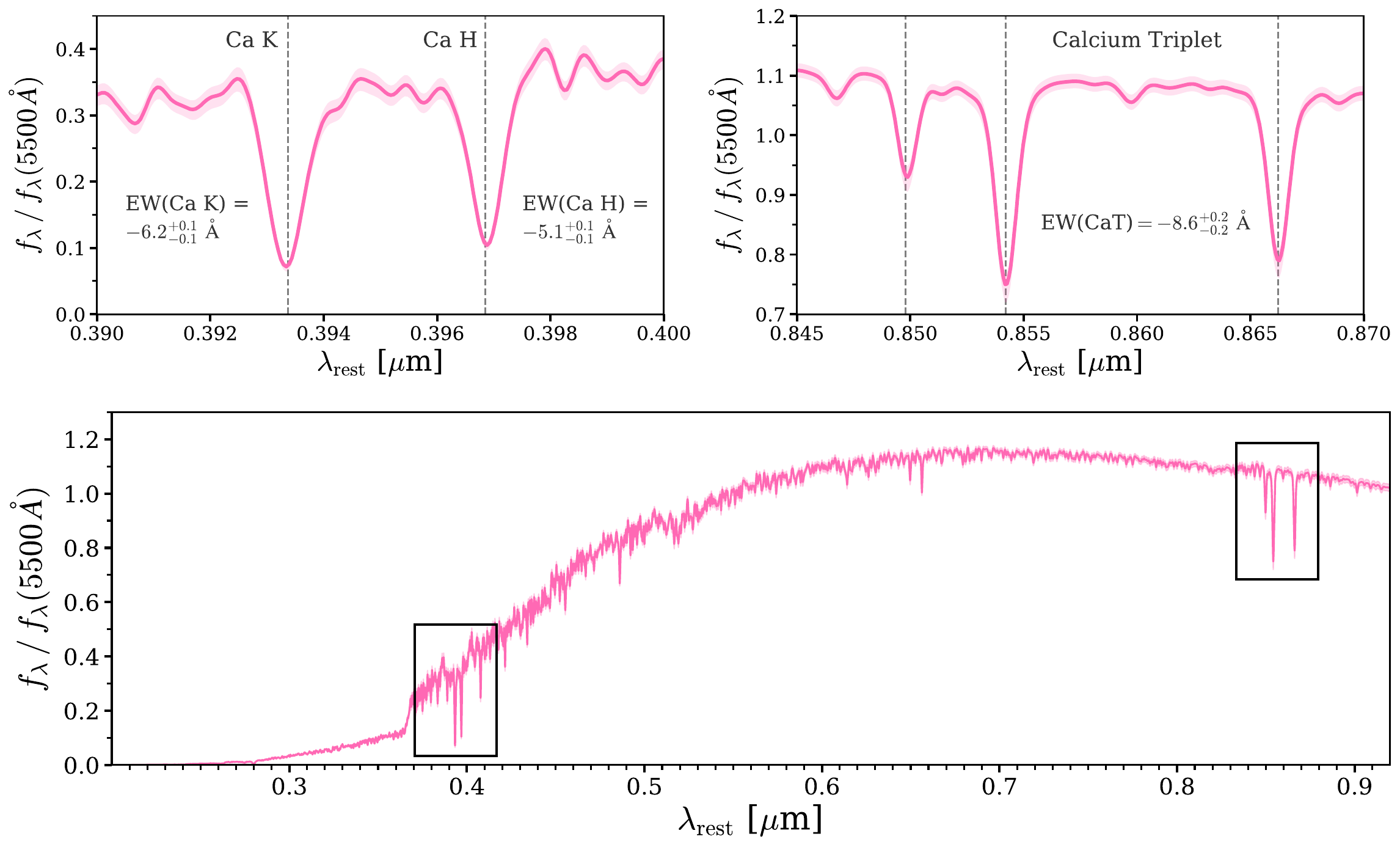}
    \caption{\textbf{Prominent Ca K, H, and triplet absorption signatures predicted for our median stack posterior.} For the stellar atmosphere fitting shown in Fig. \ref{fig:atm_fits_stack}, we predict the JWST NIRSpec H-grating spectrum for the pseudo-photosphere continuum. Note that this is not the entirety of the BH* spectrum -- we additionally expect NLTE features from the gas layers that imprint line emission and the sharp Balmer break (see \S\ref{sec:models}). (Top left) Zoomed-in view of Ca K and Ca H profiles. (Top right) Zoomed-in view of Calcium Triplet (CaT) profiles. EWs are indicated on the panels. Each line is fit with \texttt{UNITE} \citep[][]{Hviding25}; see text for details.}
\label{fig:Ca_absorption}
\end{figure*}

\subsubsection{SMS Lifetimes are Consistent with BH* Duty Cycles}

One potential concern regarding BH*s arising from SMSs is the short SMS lifetime. The typical LRD hosts spend $\sim 10$ Myrs in the LRD phase \citep[][]{Sun26}, whereas SMSs are expected to have lifetimes of $\lesssim 1-2$ Myrs \citep[e.g.,][]{Begelman10, Woods20, Nandal26SMSlifetimes}. If the LRD phase corresponds to the SMS itself, the window may be shorter still, since the SMS may only exhibit the wind structure towards the end of its life before collapsing into a black hole.

This seeming contradiction may be resolved in two ways. First, the LRD phase need not coincide with the SMS itself, but likely also includes the immediately following phase where the SMS has freshly collapsed into a BH* -- such a phase, e.g., in quasistar models, can last $\approx10-100$ Myrs \citep[e.g.,][]{Santarelli25, Begelman25, Roman-Garza26}. Our mass constraints alone cannot distinguish among the pre-collapse SMS enshrouded in dense gas and a newly formed BH* in e.g. a quasistar-like BH-plus-envelope phase \citep[e.g.,][]{Begelman08, Volonteri10, Begelman25}. Distinguishing among the different phases will require diagnostics beyond the mass that provide independent timing constraints, such as detailed chemical abundances \citep[e.g.,][]{Nandal25Nitrogen, Vasan26SPURS, Chisholm26} or the properties of the surrounding dense star cluster \citep[e.g.,][]{Inayoshi26specuni}. Irrespective of whether LRDs mark the final stages of SMSs or the birth of black holes, we are likely witnessing heavy seed formation in action.

Another factor that eases the tension with SMS lifetimes is that a single LRD could host multiple SMSs/ BH*s that may form at different times -- their lifetimes summed together would account for the \citet[][]{Sun26} host duty cycle. In support of this, a lensed galaxy at $z \sim 7$ has been found to host two BH*s separated by $\sim 70$ pc \citep[][]{Yanagisawa26}. This object demonstrates that BH* multiplicity is possible. Future observations of highly magnified LRDs could further test the prevalence of such multiplicity, as lensing resolves separations that would otherwise blend into a single point source.

\subsubsection{Binary SMS Mergers and Multiple SMSs}

Intriguingly, some of the stellar analogues that bear the most striking resemblance to BH*s are the products of binary interactions. For instance, one hypothesis for $\eta$ Carinae's Great Eruption is a stellar merger in an unstable triple system \citep[e.g.,][]{PortegiesZwart16, Smith18, Hirai21}. Similarly, luminous red novae (LRN) which share the cold $T_{\rm{eff}}$ of BH*s (Fig. \ref{fig:hr_diagram}) and Balmer line properties such as extreme decrements \citep[][]{Sneppen26LRN} are also thought to be the result of stellar mergers \citep[][]{Kaminski26rednovae}. In the case of LRN, the binary interactions may produce a dense gas envelope and a succession of slow and fast winds due to e.g. Roche lobe overflow, mass loss from the outer Lagrange point, and common envelope evolution before the merger culminates in shocks that power the spectrum \citep[e.g.,][]{Pejcha16, Metzger17, Pastorello19, Sneppen26LRN}. 

We detail these cases to illustrate that binary evolution may also be a viable channel for BH* enshrouding and formation in addition to mass loss from a single SMS. Such a binary interaction scenario may also help explain the hints of complex geometry (akin to $\eta$ Carinae; \citealt[][]{Smith18}) emerging in LRDs -- e.g., equatorial vs. polar density differences that may manifest in blueshifts in the Balmer lines of some sources versus redshifts in others \citep[e.g.,][]{Matthee26}, potential viewing angle differences that probe different column densities \citep[e.g.,][]{Madau26, Mascia26}, and polarization signatures suggestive of non-axisymmetric configurations \citep[e.g.,][]{deugenio26pol}.

We stress that our derived masses alone cannot establish that exactly a single SMS is responsible for the BH* spectrum, as the same total mass could be split among multiple SMSs, either undergoing a merger, within one cluster, or across multiple clusters in the same host galaxy that are all shining as BH*s at the same time. If there are $N$ unresolved sources with comparable properties, the fitted $T_{\rm{eff}}$ and $\log{g}$ that rely on the shape of the spectra would be unchanged, but $L_{\rm{bol}}$ which depends on the normalization is inflated by $N$ and $R_{\rm{phot}}$ by $\sqrt{N}$.

Presently, we cannot conclude whether the BH* phenomenon is powered by a single SMS, an ensemble of SMSs, or merging SMS. However, the key insight of this work is that such combinations of SMSs, including a single SMS, are now viable channels for LRD formation. Highly lensed systems \citep[e.g.,][]{Golubchik25,Yanagisawa26,Baggen25MI,Baggen26} combined with extreme adaptive optics on the ELTs may provide direct constraints on the multiplicity and binarity of SMS.

\subsection{Prediction of Observational Signatures}
\label{sec:predicted_signatures}

Our posteriors for the pseudo-photosphere continua are constrained solely by spectra at PRISM resolution. Nevertheless, the model fits allow us to predict observational signatures in higher-resolution spectra. Specifically, we take the median of the $1{,}000$ posterior draws for the median BH* stack, redshift it to $z = 4.54$ (the median redshift of the stack), convolve it with a Gaussian kernel of $\sigma \approx 120 \, \rm{km \, s}^{-1}$, and then convolve it to the resolution of the JWST H-gratings. The second convolution to match the JWST resolution is performed separately for each of the three H-gratings using its own dispersion curve $R(\lambda)$, with the nominal resolving power scaled up by a factor of $1.3$ to match the empirically measured resolution \citep[][]{Slob24}; each segment is then resampled onto an $R$-matched wavelength grid and joined at observed $1.70 \, \mu\rm{m}$ and $2.90 \, \mu\rm{m}$ to form a continuous $0.80 - 5.10 \, \mu\rm{m}$ spectrum. The resulting spectrum, shifted back to rest-frame, is displayed in Fig. \ref{fig:Ca_absorption}.

Notably, \ion{Ca}{2} 3935\AA\ (Ca K), \ion{Ca}{2} 3970\AA\ (Ca H), and the \ion{Ca}{2} 8500, 8544, 8665\AA\ triplet (CaT) absorption features stand out. Note that this predicts only one component of the BH*s we will actually observe, because as discussed previously, the atmosphere models fit the pseudo-photosphere continuum without accounting for photons encountering the clumpy (ionized + neutral), excited gas. Broadly, there are the host galaxy continuum and the BH* emission that are not accounted for in this prediction. 

We use the \texttt{UNITE} line fitting package \citep[][]{Hviding25} to measure the line EWs, and obtain EW(Ca K) $= -6.2^{+0.1}_{-0.1}$ \AA, EW(Ca H) $= -5.1^{+0.1}_{-0.1}$ \AA, and EW(CaT) $= -8.6^{+0.2}_{-0.2}$ \AA. The predicted CaT strength reflects the low $\log{g}$ of the BH*. In the hydrostatic atmosphere we assume here, low $\log{g}$ implies low pseudo-photospheric densities. As density decreases, the H$^-$ opacity is suppressed more strongly than the \ion{Ca}{2} line opacity, increasing the line-to-continuum contrast \citep[e.g.,][]{Mallik97, Cenarro02}. These \ion{Ca}{2} absorption features may serve as a key observational signature of BH*s. Indeed, CaT absorption has been observed in local LRDs such as \textit{The Egg} \citep[][]{Lin25} and is common among high-redshift LRDs (Greene et al., in prep.). In future studies, comparing the Ca velocity dispersion with that of other absorption features could offer more insight into the wind structure of LRDs. Also, the CaT strength could provide a further check on the $\log{g}$ and [M/H] we derive here, due to its sensitivity to these two parameters.

\subsection{Beyond LRDs: Implications for the Wider High-$z$ Broad-Line Population}
\label{sec:broad_line_population}

A natural question to ask now is whether the mass revision proposed here extends to the wider population of high-redshift broad-line sources. We emphasize that the mass revision here does not apply universally to all of these sources. The high-redshift broad-line population is heterogeneous, and it is useful to distinguish the following four categories.
\begin{enumerate}
    \item \textit{V-shaped LRDs:} These are the typical LRDs to which our mass revision would apply.
    
    \item \textit{Sources that host a BH*, but are not V-shaped due to host dilution:} The physics of this class of objects is the same as V-shaped LRDs, though their SEDs appear different due to the varying BH*/LRD ratios and dilution by host-galaxy light \citep[e.g.,][]{Barro25, Sun26, Rinaldi26, Perez-Gonzalez26}. Therefore, even though they may fail LRD color selections, their virial masses are likely overestimated in the same way as those of V-shaped LRDs. Note that the original V-shape selection thresholds \citep[e.g., $\beta_{\rm{opt}} > 0$ and $\beta_{\rm{UV}} < -0.37$;][]{Kocevski25stats} were chosen to build a clean sample rather than to mark a cutoff with a physical meaning. In fact, \cite{Kocevski26} recently revised the thresholds by lowering the $\beta_{\rm{opt}}$ cutoff to $-0.52$. Whether a source falls on one side or the other of such a cut therefore may have no direct implication for whether its virial mass is reliable.
    
    \item \textit{BH*s seen through thinner envelopes:} The central engine is more directly visible, most clearly through strong broad \ion{He}{2} emission, as in GN-16813 \citep[e.g.,][]{Matthee24} and GS-3073 \citep[e.g.,][]{Vanzella10NIV, Ubler23, Brazzini26}. These blue broad-line emitters, which are compact in both the UV and optical, display broad Balmer emission, but lack Balmer breaks and Balmer absorption and show high ionization lines like \ion{He}{2}. They are likely a minority: \cite{Mascia26} constrain them to $20\%$ of archival broad-line samples at $4 \leq z \leq 7$. Note that the $20\%$ may be an upper limit, as these sources are not all confirmed \ion{He}{2} emitters, and may therefore still include compact sources from category \#2. Rather than constituting a distinct population, these \ion{He}{2} emitters may extend the population of LRDs \citep[e.g.,][]{Asada26, Kocevski26} towards lower column density BH*s \citep[e.g.,][]{Matthee26, Sneppen26lbd, Mascia26, Barrufet26}. As such, their central engines are essentially similar to those of LRDs -- indeed, they are X-ray faint \citep[e.g.,][]{Brazzini26} and show little variability \citep[e.g.,][]{Liu26TWINKLE} unlike classical AGN. Therefore, their masses based on virial estimators and the interpretation of broad Balmer lines may also be unreliable -- indeed, some phases of Type IIn SNe show Balmer line spectra mirroring these blue LRDs while hosting no kinematic broad-line region whatsoever \citep[][]{Naidu26}. However, the methods calibrated here relying on pseudo-photospheric continuua and absorber velocities may be inapplicable to such lower column density sources given the absence of these features. We note that sources in categories \#2 and \#3 are selected as ``Little Blue Dots'' as per the \citet[][]{Geris26} criteria (see also \citealt[][]{Madau26}).
    
    \item \textit{Traditional AGN and quasars:} These are X-ray detected, show hot dust emission, display variability, and lack the Balmer absorption and Balmer breaks ubiquitous in LRDs \citep[e.g.,][]{Yang23ASPIRE, Bulichi26, Yang26EUCLID}. Crucially, their spectra look exactly like those of their low-redshift counterparts, so local calibrations may be applied somewhat reliably \citep[e.g.,][]{Abuter24}. Luminous quasars \citep[e.g.,][]{Liu25quasars, Silverman25quasars, Marshall25quasars} and dust-reddened AGN such as GNz7q \citep[][]{Fujimoto22, Fei26} are inferred to have black hole masses of $\sim 10^{7-9} \, M_\odot$, and extend the LRD range we bracket in Table \ref{tab:mass_upper_bounds} to higher masses.
\end{enumerate}

The applicability of virial mass estimates calibrated on $z\approx0$ sources therefore follows the physics rather than the LRD color selection. Based on the consideration discussed here, virial estimators may provide robust estimates only for category \#4 whereas categories \#1, \#2, and \#3 may require novel mass estimators like those discussed here.

\section{Summary}
\label{sec:summary}

We explore emerging approaches to estimating the masses of LRD central engines (BH*s) and, from the resulting low masses, argue that BH*s may arise from single SMSs.

\vspace{0.2cm}
\noindent $\bigstar$ We first characterize the BH* pseudo-photospheres, whose parameters allow us to derive mass estimates. Specifically:

\begin{itemize}
    \item To estimate BH* masses without contamination from their host galaxies, we first subtract the hosts to isolate the BH* spectra, following \cite{Sun26}. Splitting the BH*s into three luminosity substacks reveals a diversity in spectral shape -- luminous BH*s peak at bluer wavelengths and are UV-bright. [\S\ref{sec:lrd_decomposition}; Figs. \ref{fig:substack_gallery}, \ref{fig:substacks}]

    \item We then fit the BH* pseudo-photosphere continua with optically thick atmosphere models from \cite{Liu26TLUSTY}, which extend to far lower gravities than standard stellar libraries. These models describe the pseudo-photosphere without accounting for the gas layers that imprint features such as the Balmer break, emission lines, and Balmer absorption, so our fits focus on the optical-to-near-IR continuum redward of the Balmer break. [\S\ref{sec:models}, \S\ref{sec:model_fitting}; Figs. \ref{fig:atm_fits_stack}, \ref{fig:vary_parameter}, \ref{fig:atm_fits_substacks}]
    
    \item The fits return a self-consistent set of pseudo-photospheric parameters, including $T_{\rm{eff}}$, $L_{\rm{bol}}$, $\log{g}$, and [M/H]. By simultaneously constraining temperature, density, and metallicity rather than absorbing all SED variation into a single temperature, these fits place the BH* substacks within a much narrower range of $T_{\rm{eff}}$ than modified blackbody fits do. Together, the $T_{\rm{eff}}$ and $L_{\rm{bol}}$ place BH*s in the coldest yet most luminous corner of the HR diagram. [\S\ref{sec:parameters}; Table \ref{tab:substacks}, Fig. \ref{fig:hr_diagram}]
\end{itemize}

\noindent $\bigstar$ Using the pseudo-photospheric parameters from the model fits, we estimate the BH* mass with four different methods, following \cite{Naidu26}.

\begin{itemize}
    \item \textit{Surface gravity}: The model fits return $\log{g} \approx -2.5 - -1.9$, much lower than even the gravities of hypergiants. Combined with the pseudo-photospheric radius $R_{\rm{phot}}$, the $\log{g}$ sets the mass. [\S\ref{sec:surface_gravity}]

    \item \textit{$L_{\rm{bol}}/L_{\rm{Edd}}$}: If BH*s radiate above their Eddington luminosities like the enshrouded eruptions they resemble, then their $L_{\rm{bol}} / L_{\rm{Edd}}$ sets the mass. [\S\ref{sec:eddington_lum}]

    \item \textit{Escape velocity}: In an optically thick massive-star wind, the outflow reaches its terminal velocity $v_\infty$ below the pseudo-photosphere, so $v_\infty$ exceeds the local escape velocity $v_{\rm{esc}}$ at the pseudo-photosphere. Approximating $v_\infty$ with the blue edge of the absorption trough bounds the mass from above. [\S\ref{sec:escape_velocity}]

    \item \textit{Variability}: A pseudo-photosphere can respond to a perturbation over its dynamical time, $t_{\rm{dyn}} = R_{\rm{phot}}/v_{\rm{esc}}$. A more massive BH* binds its envelope more tightly, so the envelope responds sooner. The lack of variability thus caps how massive the BH* can be. [\S\ref{sec:variability}]
\end{itemize}

\noindent $\bigstar$ Under the pseudo-photosphere picture, all four methods yield consistent fiducial mass estimates (Tables \ref{tab:mass_upper_bounds}, \ref{tab:mass}, Figs. \ref{fig:virial_comparison}, \ref{fig:m_vs_mstar}). For the typical BH*, they converge on $\approx 10^{4-5} \, M_\odot$. Across the three substacks, the masses span $\approx 10^{3.5-6} \, M_\odot$. This carries several implications:

\begin{itemize}
    \item \textit{BH*s are not ``overmassive''}: Measured against host stellar masses of $\approx 10^8 \, M_\odot$, our BH* masses are $\sim 3$ dex lower than estimates that report black holes orders of magnitude above the local scaling relation between black hole mass and stellar mass. [\S\ref{sec:mass_summary}; Figs. \ref{fig:virial_comparison}, \ref{fig:m_vs_mstar}]

    \item \textit{BH* masses are consistent with an SMS origin}: All the derived BH* masses fall below the mass ceiling of SMSs imposed by general relativistic instabilities, $\approx 10^{5-6} \, M_\odot$, suggesting that a BH* is consistent with arising from a single SMS. The low metallicities from the model fits, [M/H] $\approx -1.9 - -1.8$, place BH*s in the metal-poor environments where SMSs are expected to form. [\S\ref{sec:mass_discussion}; Fig. \ref{fig:m_vs_mstar}]

    \item \textit{BH*s have super-Eddington luminosities and explain the LRD LF bright end cutoff}: Reconciling the low BH* masses with the $L_{\rm{bol}}$ obtained from the model fits requires $L_{\rm{bol}}/L_{\rm{Edd}} \sim 50$. At this ratio, a maximal SMS would reach $L_{\rm{bol}} \sim 10^{44.8-45.8} \, \rm{erg\ s}^{-1}$, which coincides with the bright-end cutoff of the observed LRD luminosity function at $\approx 10^{45-46} \, \rm{erg\ s}^{-1}$. [\S\ref{sec:mass_discussion}]
\end{itemize}

\noindent $\bigstar$ \textit{BH*s are predicted to display Ca absorption:} We predict BH* features at higher resolution based on our fits to the PRISM spectra. The low surface gravity predicts strong Ca H, K, and triplet absorption. [\S\ref{sec:predicted_signatures}; Fig. \ref{fig:Ca_absorption}]

\vspace{0.3cm}
\noindent $\bigstar$ \textit{Beyond LRDs}:  We contextualize our proposed mass revision among the wider high-$z$ broad-line population. The mass revision would apply to V-shaped LRDs, sources that host a BH* but are not V-shaped due to host dilution, and BH*s seen through thinner envelopes. Virial estimators likely apply to quasars and objects that resemble classical AGN. [\S\ref{sec:broad_line_population}]

\vspace{0.2cm}
Our derived parameters, including the low BH* masses, make an SMS origin a possible and promising channel for LRD formation, one that had previously been ruled out by virial masses. If this scenario is, as we suggest, a scaled-up version of the familiar massive star phenomena, we may be witnessing heavy seed formation in action.

\section*{Acknowledgments} We acknowledge insightful conversations with Luc Dessart, Sandro Tacchella, and Marta Volonteri. WQS and RPN acknowledge funding from {\it JWST} programs GO-3516, GO-5224, and the MIT Undergraduate Research Opportunities Program (UROP). Support for this work was provided by NASA through the NASA Hubble Fellowship grant HST-HF2-51515.001-A awarded by the Space Telescope Science Institute, which is operated by the Association of Universities for Research in Astronomy, Incorporated, under NASA contract NAS5-26555. RPN thanks Neil Pappalardo and Jane Pappalardo for their generous support of the MIT Pappalardo Fellowships in Physics, and for their enthusiasm and encouragement for pursuing the earliest galaxies and black holes. JM and AT acknowledge funding from the European Union (ERC, AGENTS, 101076224). KEH acknowledges support from the Independent Research Fund Denmark (DFF) under grant 5251-00009B and co-funding by the European Union (ERC, HEAVYMETAL, 101071865). Views and opinions expressed are, however, those of the authors only and do not necessarily reflect those of the European Union or the European Research Council. Neither the European Union nor the granting authority can be held responsible for them. D.H. is supported by STScI grants HST-GO17770.002, JWST-GO-12468.001, and JWST-GO09964.001. P.N. acknowledges support from the Gordon and Betty Moore Foundation and the John Templeton Foundation, which fund the Black Hole Initiative (BHI) at Harvard University, where she is a PI. P.N. also acknowledges support from STScI/NASA via grant JWST-GO-03293024.

The data products presented herein were retrieved from the Dawn JWST Archive (DJA). DJA is an initiative of the Cosmic Dawn Center (DAWN), which is funded by the Danish National Research Foundation under grant DNRF140. This work is based on observations made with the NASA/ESA/CSA James Webb Space Telescope. The data were obtained from the
Mikulski Archive for Space Telescopes at the Space Telescope Science Institute, which is operated by the Association of Universities for Research in Astronomy, Inc., under NASA contract NAS 5-03127
for JWST. Support for programs
\#3516, \#5224, \#5664 was provided by NASA through grants from the Space
Telescope Science Institute, which is operated by the Association of
Universities for Research in Astronomy, Inc., under NASA contract
NAS 5-03127. 

The spectra used in this paper are associated with programs 1180 \citep{DEugenio25}, 1181 (PI: D. Eisenstein), 1199 \citep{Stiavelli23}, 1207 (PI: G. Rieke), 1208 \citep{Willott22}, 1210 (PI: N. Luetzgendorf), 1211 \citep{Maseda24}, 1212 - 1215 (PI: N. Luetzgendorf), 1219 (PI: N. Luetzgendorf), 1222 \citep{Christensen23}, 1228 \citep{Luhman24}, 1229 \citep{Luhman24b}, 1286 (PI: N. Luetzgendorf), 1287 (PI: K. Isaak), 1324 \citep{Mascia24}, 1345 \citep{Finkelstein23}, 1433 \citep{Hsiao24}, 1635 (PI: C. Martin), 1671 \citep{Maseda23}, 1747 (PI: G. Roberts-Borsani), 1810 \citep{Belli24}, 1835 \citep{Wang25}, 1869 (PI: D. Schaerer), 1871 \citep{Chisholm24}, 1879 \citep{Cataldi25}, 1914 \citep{Shapley25}, 2028 \citep{Wang24j0910}, 2073 (PI: J. Hennawi), 2110 \citep{Slob24}, 2198 \citep{Barrufet25}, 2282 \citep{Bradley23}, 2478 \citep{Topping25}, 2561 \citep{Bezanson24}, 2565 \citep{Nanayakkara25}, 2593 \citep{Strom23}, 2640 (PI: W. Best), 2674 (PI: P. Arrabal Haro), 2736 \citep{Pontoppidan22}, 2750 \citep{ArrabalHaro23}, 2756 \citep{Mascia24}, 2758 (PI: M. Stiavelli), 2767 \citep{Williams23rxj}, 2770 (PI: M. McCaughrean), 3073 \citep{Castellano24abell}, 3117 \citep{Yue25}, 3215 \citep{Eisenstein25}, 3222 \citep{Scibelli25}, 3325 (PI: F. Wang), 3503 (PI: A. Adamo), 3543 \citep{Carnall24}, 3567 \citep{Ito26}, 3788 (PI: D. Weisz), 4106 (PI: E. Nelson), 4212 \citep{Bradley25}, 4233 \citep{degraaff25rubies}, 4246 \citep{Abdurrouf24}, 4265 \citep{Solimano25}, 4287 \citep{Tang25}, 4318 (PI: J. Antwi-Danso), 4446 \citep{Frye24}, 4557 (PI: H. Yan), 4750 \citep{Nakajima26}, 4762 (PI: S. Fujimoto), 5105 \citep{Shen24nexus}, 5224 (PIs: P.A. Oesch \& R.P. Naidu), 6154 (PI: A. de Graaff), 6368 (PI: M. Dickinson), 6541 \citep{DeCoursey25}, 6585 (PI: D. Coulter), 6642 (PI: J. Muzerolle Page), 9223 (PI: S. Fujimoto), and FRESCO IFU \citep{Matthee24, Torralba25IFU}.

Software used in developing this work includes: \texttt{dynesty} \citep{Speagle19}, \texttt{UNITE} \citep{Hviding25}, \texttt{msaexp} \citep{msaexp}, \texttt{Prospector} \citep{Leja17, Leja19, Johnson21}, \texttt{lmfit} \citep{lmfit}, \texttt{SciPy} \citep{scipy}, \texttt{dust\_extinction} \citep{Gordon_dust_models}, \texttt{Astropy} \citep{astropy}, \texttt{numpy} \citep{numpy}, \texttt{matplotlib} \citep{matplotlib}, \texttt{jupyter} \citep{jupyter}, \texttt{IPython} \citep{ipython}, \texttt{claude} \citep{Anthropic2026Claude}.

During the preparation of this manuscript, the authors used Claude (Fable 5, Opus 5; Anthropic) to assist with developing and cleaning up code and checking for grammar mistakes. All AI-assisted output was reviewed and verified by the authors, who take full responsibility for the content of this work.

\bibliography{MasterBiblio}
\bibliographystyle{apj}

\appendix

\section{Posteriors for the BH* Substacks}

\begin{figure*}
    \centering
    \includegraphics[width=\linewidth]{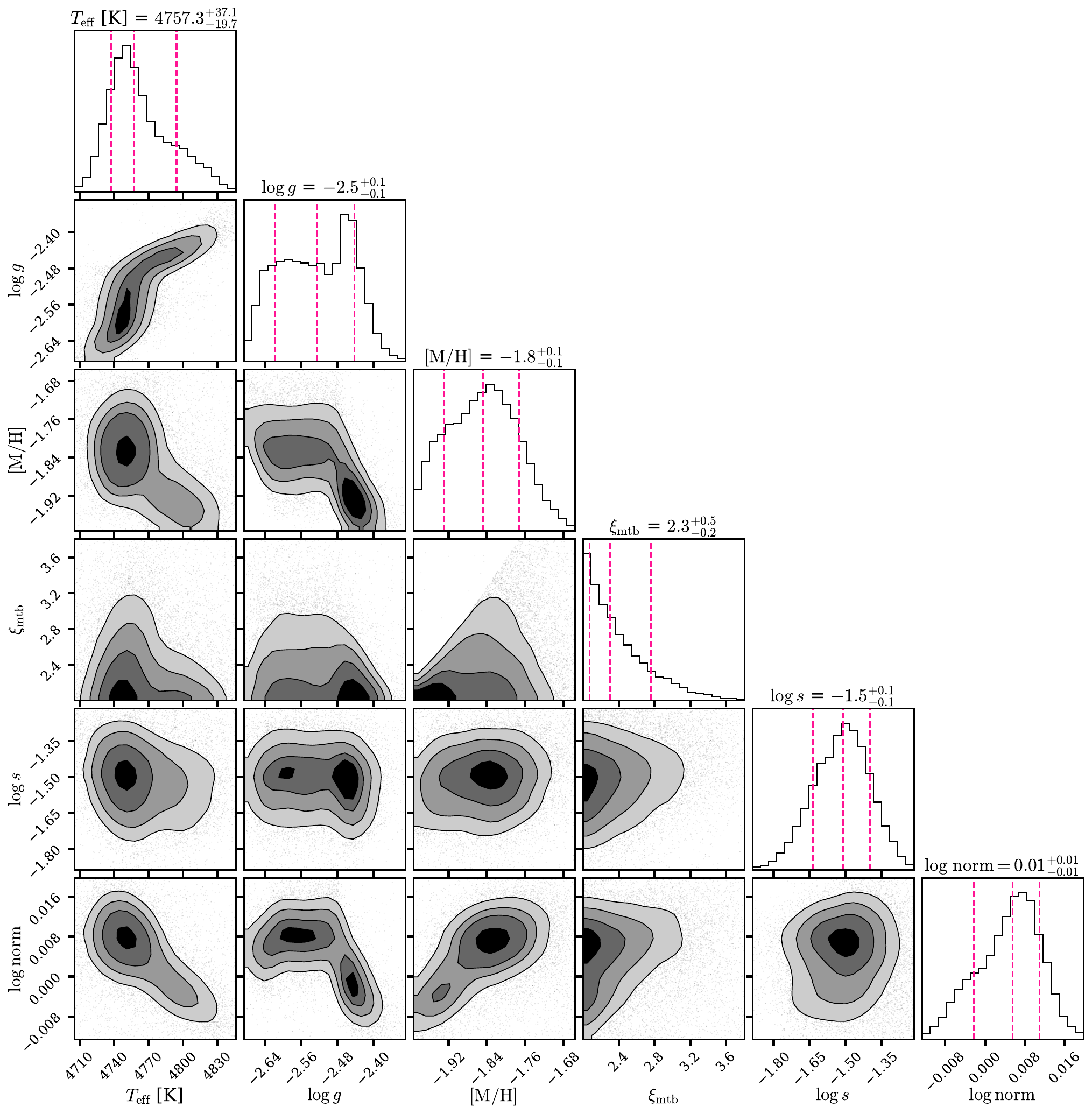}
    \caption{Following Fig. \ref{fig:atm_fits_stack}, for the \textbf{luminous} BH* substack.}
\label{fig:corner_luminous_substack}
\end{figure*}

\begin{figure*}
    \centering
    \includegraphics[width=\linewidth]{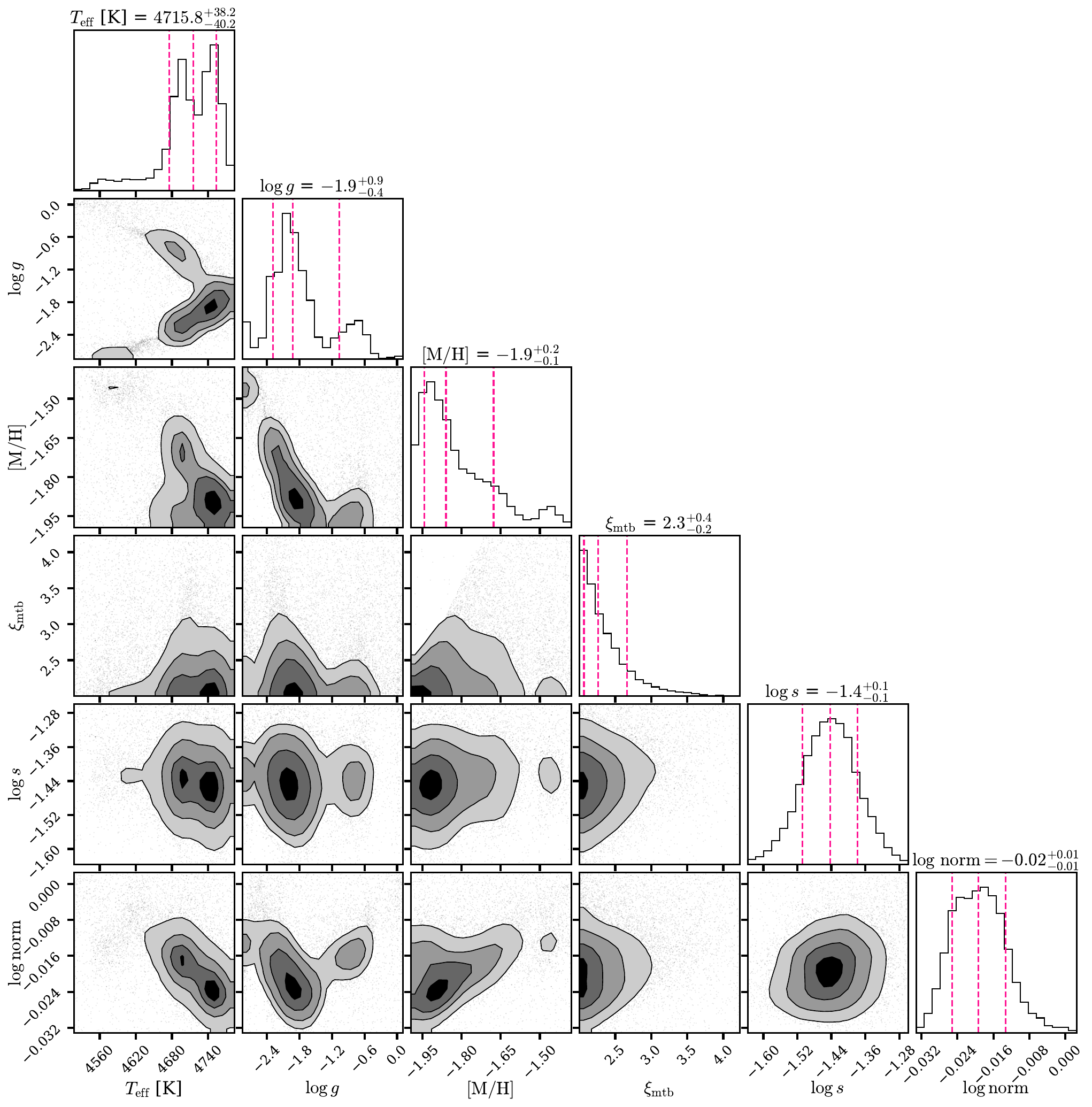}
    \caption{Following Fig. \ref{fig:atm_fits_stack}, for the \textbf{intermediate} BH* substack.}
\label{fig:corner_inter_substack}
\end{figure*}

\begin{figure*}
    \centering
    \includegraphics[width=\linewidth]{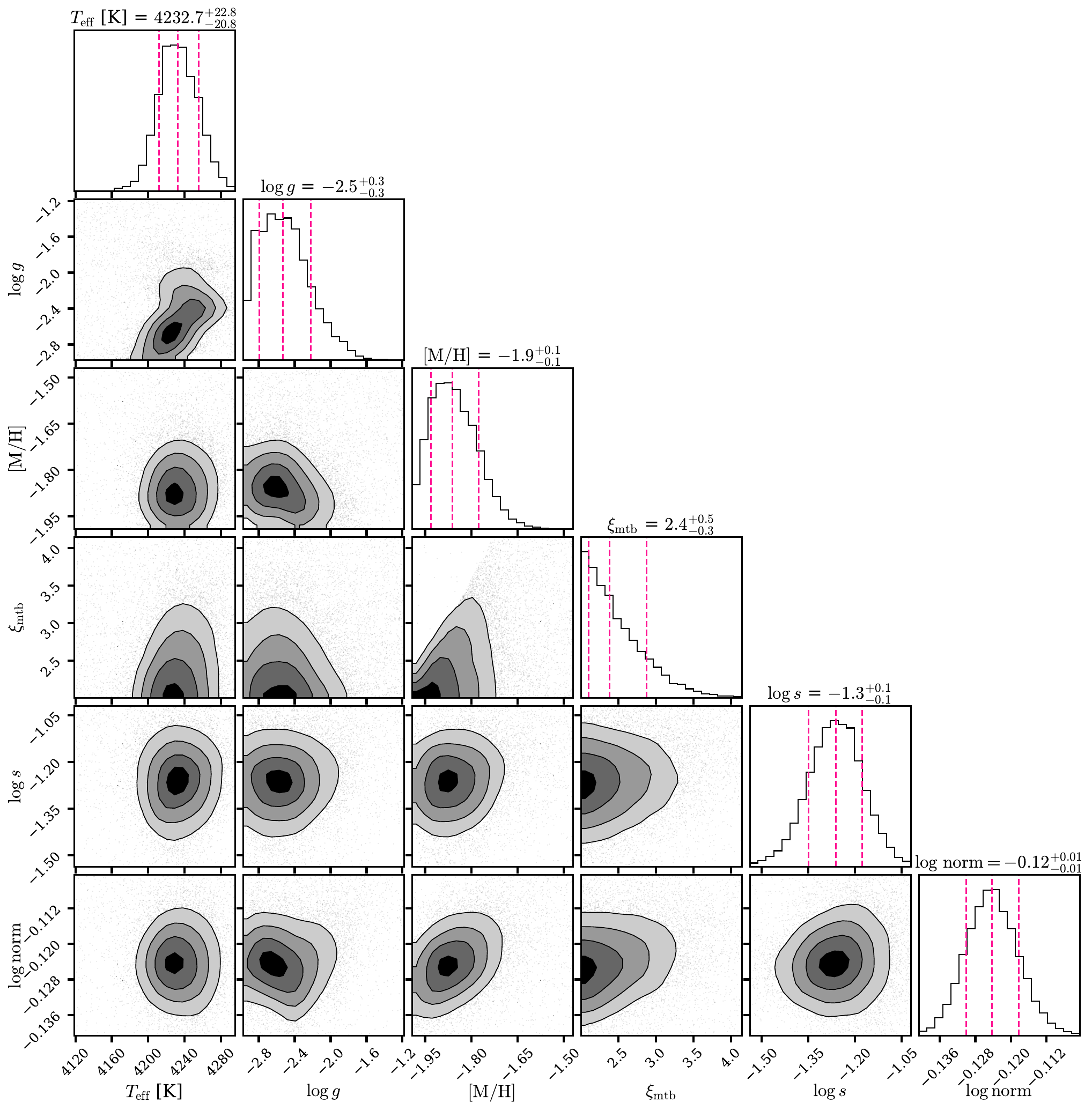}
    \caption{Following Fig. \ref{fig:atm_fits_stack}, for the \textbf{faint} BH* substack.}
\label{fig:corner_faint_substack}
\end{figure*}

Figs. \ref{fig:corner_luminous_substack}, \ref{fig:corner_inter_substack}, \ref{fig:corner_faint_substack} show the corner plots of the \texttt{dynesty} posteriors for the three BH* substacks, following Fig. \ref{fig:atm_fits_stack}.

\section{\texttt{Prospector} Modeling for Host Galaxies}

\begin{figure*}
    \centering
    \includegraphics[width=0.80\linewidth]{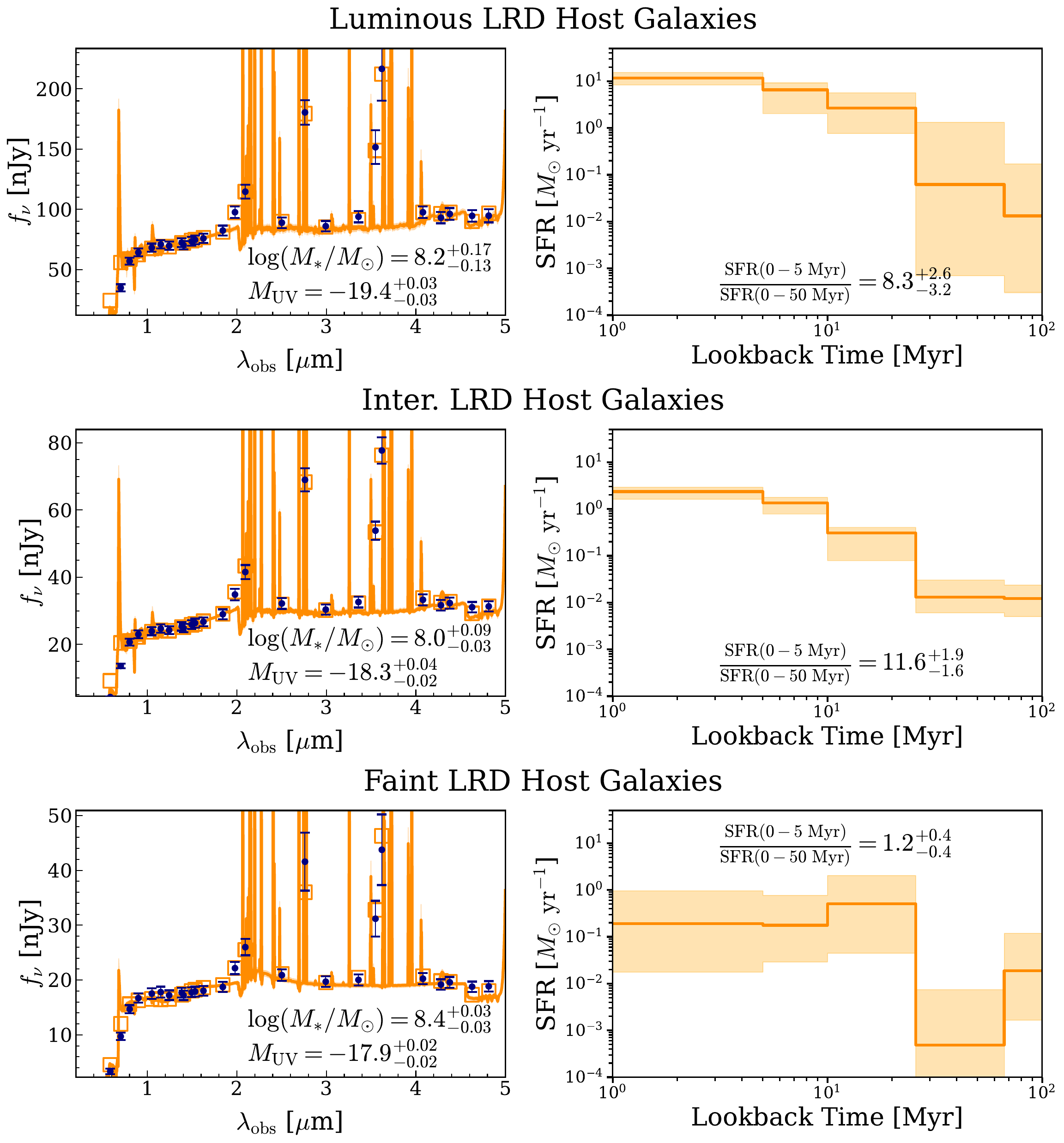}
    \caption{\textbf{Stellar masses of LRD host galaxies are derived from \texttt{Prospector} modeling} \citep[][]{Leja17, Leja19, Johnson21}. Posterior draws in orange are based on \texttt{Prospector} fits to the median host galaxy stacks. The stacks are fit in the form of synthesized JWST NIRCam photometry in all medium- and broad-bands (navy) along with fluxes of strong lines (H$\alpha$, H$\beta$, [\ion{O}{3}]).}
\label{fig:host_substacks_prospector}
\end{figure*}

Fig. \ref{fig:host_substacks_prospector} shows the \texttt{Prospector} SED fits from which we estimate LRD host stellar masses reported in Table \ref{tab:substacks} and Figs. \ref{fig:virial_comparison}, \ref{fig:m_vs_mstar}.

\end{document}